\documentclass[%
prx,%
reprint,%
amsmath,amssymb,%
floatfix,%
footinbib,%
superscriptaddress
]{revtex4-2}

\usepackage[T1]{fontenc}
\usepackage[latin9]{inputenc}
\usepackage[english]{babel}
\usepackage{mathrsfs}
\usepackage{bbm}
\usepackage{graphicx}
\usepackage{xcolor}
\usepackage{physics}
\usepackage{float}
\usepackage{lipsum}
\usepackage{enumerate}
\usepackage{bm}

\usepackage[colorlinks,
            linkcolor=blue,    
            anchorcolor=black,
            citecolor=blue,  
            urlcolor=blue,    
 	        bookmarks=true,        
	        bookmarksopen=true,    
	        bookmarksnumbered=true,
]{hyperref}

\global\long\def\kket#1{\ket{\ket{#1}}}

\global\long\def\av#1{\left\langle #1 \right\rangle }

\global\long\def\abs#1{\left|#1\right|}
\renewcommand*\d{\mathop{}\!\mathrm{d}}
\newcommand{\dirac}[1]{\,\delta\!\left(#1\right)}
\newcommand{\heav}[1]{\,\Theta\!\left(#1\right)}
\newcommand{\s}{\mathfrak{s}}
\renewcommand{\vec}[1]{\bm{#1}}

\begin{document}

\title{Complex Spacing Ratios: A Signature of Dissipative Quantum Chaos}

\author{Lucas  S\'a}
\email{lucas.seara.sa@tecnico.ulisboa.pt}

\affiliation{CeFEMA, Instituto Superior T\'ecnico, Universidade de Lisboa, Av.\ Rovisco Pais, 1049-001 Lisboa, Portugal}

\author{Pedro Ribeiro}
\email{ribeiro.pedro@tecnico.ulisboa.pt}

\affiliation{CeFEMA, Instituto Superior T\'ecnico, Universidade de Lisboa, Av.\ Rovisco Pais, 1049-001 Lisboa, Portugal}
\affiliation{Beijing Computational Science Research Center, Beijing 100193, China}

\author{Toma\v z Prosen}
\email{tomaz.prosen@fmf.uni-lj.si}

\affiliation{Department of Physics, Faculty of Mathematics and Physics, University of Ljubljana, Ljubljana, Slovenia}

\begin{abstract}
We introduce a complex-plane generalization of the consecutive level-spacing ratio distribution used to distinguish regular from chaotic quantum spectra. 
Our approach features the distribution of complex-valued ratios between nearest- and next-to-nearest-neighbor spacings. We show that this quantity can successfully detect the chaotic or regular nature of complex-valued spectra, which is done in two steps. 
First, we show that, if eigenvalues are uncorrelated, the distribution of complex spacing ratios is flat within the unit circle, whereas random matrices show a strong angular dependence in addition to the usual level repulsion. The universal fluctuations of Gaussian Unitary and Ginibre Unitary universality classes in the large-matrix-size limit are shown to be well described by Wigner-like surmises for small-size matrices with eigenvalues on the circle and on the two-torus, respectively. To study the latter case, we introduce the Toric Unitary Ensemble, characterized by a flat joint eigenvalue distribution on the two-torus. 
Second, we study different physical situations where non-Hermitian matrices arise: dissipative quantum systems described by a Lindbladian, nonunitary quantum dynamics described by non-Hermitian Hamiltonians, and classical stochastic processes. We show that known integrable models have a flat distribution of complex spacing ratios whereas generic cases, expected to be chaotic, conform to Random Matrix Theory predictions. Specifically, we are able to clearly distinguish chaotic from integrable dynamics in boundary-driven dissipative spin-chain Liouvillians and in the classical asymmetric simple exclusion process and to differentiate localized from delocalized regimes in a non-Hermitian disordered many-body system.
\end{abstract}

\maketitle

\section{Introduction}
Understanding decoherence and dissipation effects arising in open quantum mechanical systems requires dealing with nonunitary dynamics generated by non-Hermitian operators. 
Non-Hermitian physics has attracted much attention recently, for instance, in the study of Lindbladian dynamics of integrable~\cite{prosen2008,prosen2010,ribeiro2019,rowlands2018,medvedyeva2016,eisler2011,banchi2017,prosen2011,prosen2015,karevski2013,ilievski2017} and chaotic~\cite{sa2019,denisov2018,can2019a,can2019} open quantum systems, topological phases of open systems~\cite{zeuner2015,gong2018,harari2018,bandres2018,shen2018,yao2018,lee2016,kawabata2018A,leykam2017,kawabata2018ST,kawabata2018PTSC,kawabata2019}, $\mathscr{P\!T}$-symmetric and general non-Hermitian optics~\cite{makris2008,klaiman2008,guo2009,ruter2010,konotop2016,elganainy2018,feng2017,miri2019}, non-Hermitian many-body localization~\cite{hamazaki2018}, non-Hermitian quantum critical phenomena~\cite{lee2014A,lee2014B,ashida2017,wei2017}, or quantum chaotic scattering~\cite{schomerus2017,huang2018}. However, a methodology to classify all of these non-Hermitian systems into different classes or phases, in terms of their universal spectral correlations, is still lacking.

For Hermitian systems, the by-now universally accepted conjectures of Berry and Tabor~\cite{berry1977} and of Bohigas, Giannoni, and Schmit~\cite{bohigas1984} (see also Ref.~\cite{casati1980}) assert, respectively, that classically integrable systems follow Poisson statistics of uncorrelated random variables, while systems with a chaotic semiclassical limit have statistics well described by Random Matrix Theory (RMT). Most astonishingly, many-body systems with no classical counterpart follow a similar rule. Poisson level statistics is found for integrable or (many-body) localized systems whereas RMT distributions are observed in generic thermalizing phases~\cite{rigol2008,nandkishore2015}. The power of the RMT approach relies on the fact that spectral fluctuations (measuring correlations of levels) are highly universal, depending solely on the symmetries of the system, and not on the details of particular models. For instance, the three classical Gaussian ensembles (GOE, GUE, and GSE) are completely determined by time-reversal symmetry, depending on a single parameter $\beta=1$, $2$, or $4$, the Dyson index.

Since the early days of RMT, level-spacing distributions, i.e.\ the distribution of the distance, $s=\varepsilon_{i+1}-\varepsilon_{i}>0$, between consecutive energy levels, $\varepsilon_{i+1},\varepsilon_{i}$, have proved to be a very useful and hence popular measure of spectral correlations in integrable and chaotic systems, i.e.\ a \emph{signature of quantum chaos}. Indeed, for closed systems, spacings between uncorrelated levels display level clustering, while RMT statistics lead to level repulsion, with a characteristic power-law behavior of the spacing distribution, $P(s)\propto s^\beta$ as $s\to0$, in the respective universality classes. Rather remarkably, the spacing distribution in the (universal) large-matrix-size limit is very well described by that obtained for $2\times2$ matrices, the Wigner surmise. Spacing distributions further allow the study of intermediate statistics, either with crossovers between Poisson and RMT statistics~\cite{brody1973,berry1984,lenz1991b,lenz1991c,prosen1993,prosen1994,bogomolny1999,bogomolny2001} or transitions between different RMT universality classes~\cite{mehta1983,pandey1983,lenz1991a,lenz1991b,schierenberg2012}. Statistics of higher-order spacings (i.e.\ distance between $k$th-nearest neighbors) have also been considered over the years~\cite{dyson1962iii,gunson1962,sakhr2006,srivastava2018,abul-magd1999,abul-magd2000}.

For non-Hermitian systems, by a direct generalization~\cite{grobe1988} of the Berry-Tabor and Bohigas-Giannoni-Schmit conjectures to dissipative systems, we expect classically integrable systems and classically chaotic systems to follow Poisson and Ginibre level statistics, respectively. For random matrices from the Ginibre ensembles (i.e.\ matrices where all entries are independent and identically distributed (iid) Gaussian random variables) one finds cubic level repulsion, $P(s)\propto s^3$. Interestingly, all three Ginibre ensembles (GinOE, GinUE, and GinSE) have the same cubic level repulsion~\cite{grobe1988,grobe1989,haake2013,akemann2019}, independently of the Dyson index $\beta$. For those ensembles, a Wigner-like surmise, in terms of modified Bessel functions, has recently been proposed in Ref.~\cite{hamazaki2019}, in which it was also shown that noncubic level repulsion can exist in non-Hermitian ensembles with different symmetries.

In order to compare theoretical predictions of RMT with actual measured or computed level sequences, one has to eliminate the dependence of the spacing distribution on the local mean spectral density, which is nonuniversal and system-dependent. This elimination is achieved by a procedure known as \emph{unfolding}~\cite{haake2013,guhr1998}, in which, in the case of a real spectrum, one changes from a sequence $E_j$ of levels to a new sequence $e_j=\mathcal{N}(E_j)$, where $\mathcal{N}(x)$ is the level staircase function measuring the mean number of levels below $x$. At the unfolded scale, the spacing distribution has a mean unit spacing and thus fluctuations can be uniformly compared across the spectrum.
Unfolding is a nontrivial procedure since it requires an analytic expression (or accurate estimate) of the level density, which is not available in general. Furthermore, numerical unfolding sometimes proves ambiguous and numerically unreliable. In the case of a two-dimensional---i.e.\ complex---spectrum the situation is worse: there the unfolding is, in principle, ambiguous; even so, one can find a minimal prescription that guarantees uniform unfolded complex level density~\cite{akemann2019}.

An alternative way to overcome the local dependence on the level density is to consider ratios of consecutive spacings, which were introduced in Ref.~\cite{oganesyan2007}. They were extensively applied in numerical studies of many-body localization~\cite{oganesyan2007,pal2010,cuevas2012,iyer2013,laumann2014,luitz2015,johri2015,argawal2015,chen2018,buijsman2019}, periodically driven, interacting quantum systems~\cite{d'alessio2014}, and quantum quenches~\cite{kollath2010,collura2012}. In Refs.~\cite{atas2013,atas2013long}, analytic expressions for the ratio distributions were obtained, including Wigner-like surmises for $3\times3$ matrices. The transition between Poisson and GOE statistics at the level of ratios~\cite{chavda2013}, higher-order spacing ratios~\cite{atas2013long,tekur2018pre,tekur2018prb,tekur2018c,bhosale2018} and nearest-neighbor by next-to-nearest-neighbor ratios (NN-by-NNN ratios)~\cite{srivastava2018} have also been considered recently.

While spacing (and spacing ratio) distributions for real spectra are well understood~\cite{guhr1998,mehta2004,haake2013,forrester2010,forrester2004}, and some results exist for spacings in complex spectra~\cite{grobe1988,grobe1989,akemann2009,haake2013,fyodorov1997,hamazaki2019,akemann2019}, two major shortcomings in the latter case remain to be addressed. On the one hand, to bypass the difficult and unreliable unfolding procedure, one is naturally led to consider ratios of spacings in the complex plane; However, this issue remains an open question. On the other hand, the existing studies on spacings in complex spectra focused solely on the distance, $s > 0$, between the complex eigenvalue and its nearest neighbor, neglecting the additional information contained in the 
angular (directional) correlations.

In this paper, we tackle both issues above by introducing \emph{complex spacing ratios}, as the ratio of the distance (taken as a complex number with magnitude and direction) from a given level to its nearest neighbor (NN) by the (complex) distance to the next-to-nearest neighbor (NNN); for a precise definition see Sec.~\ref{sec:main_results}. Two comments are in order regarding these complex spacing ratios. First, when defining ratios for real spectra, level sequences are usually assumed to be ordered. However, there is no global order in the complex plane, and hence all ratios that relied on the ordering have to be abandoned. Indeed, the only remaining spacing ratio is the NN-by-NNN ratio, the modulus of which was introduced in Ref.~\cite{srivastava2018} (and $k$th-nearest neighbor generalizations) for studies of real spectra. Second, this new spacing ratio (and not only its modulus!) can also be defined for real spectra. It does \emph{not} coincide with \emph{any} of the aforementioned ratios; in particular, it adds a sign to the NN-by-NNN ratio of Ref.~\cite{srivastava2018}. We emphasize that, while this sign might seem a minor difference in the case of real spectra, for complex spectra, the full angular dependence constitutes, arguably, the cleanest signature of dissipative quantum chaos.

The paper is organized as follows. In Sec.~\ref{sec:main_results} we define the complex spacing ratio, mention some of its qualitative features, point out the differences for integrable and chaotic spectra and state the key ideas behind our analytical results. In Sec.~\ref{sec:Surmises} we present exact analytical distributions and small-$N$ surmises. In Sec.~\ref{sec:examples} examples of application to different physical problems (driven spin-chains, non-Hermitian many-body localization, and classical stochastic processes) are studied. We draw our conclusions in Sec.~\ref{sec:conclusions}. A detailed derivation of the analytical results is given in three appendices: the ratio distributions for uncorrelated random variables in $d$ dimensions are computed in Appendix~\ref{sec:poisson}; exact analytical distributions and small-size surmises are derived for Hermitian random matrix ensembles in Appendix~\ref{sec:analytics_real}, and for non-Hermitian ensembles in Appendix~\ref{sec:analytics_complex}.

\section{Overview and main results}
\label{sec:main_results}

\begin{figure}[tbp]
    \centering
    \includegraphics[width=0.99\columnwidth]{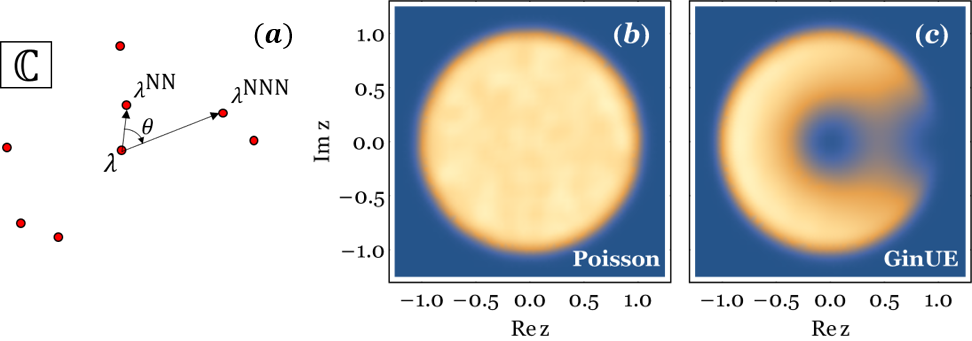}
    \caption{$(a)$: sketch of the NN and NNN level spacings used to define the complex spacing ratio, $z$. $(b)$ and $(c)$:  density plot of $z$ in complex plane for $(b)$ $10^5$ uncorrelated levels, and $(c)$ $100$
    $N \times N$ random matrices drawn from the GinUE with $N=10^4$.}
    \label{fig:Poisson_vs_GinUE}
\end{figure}

Let the set $\{\lambda_k\}_{k=1}^N$ be the spectrum of some Hermitian or non-Hermitian matrix.
The levels $\lambda_k$ may, correspondingly, be real or complex. For each $\lambda_k$, we find its NN (with respect to the distance in $\mathbb R$ or in $\mathbb C$), $\lambda_k^\mathrm{NN}$, and its NNN, $\lambda_k^\mathrm{NNN}$, and define the (in general complex) ratio 
\begin{equation}
    z_k=\frac{\lambda_k^\mathrm{NN}-\lambda_k}{\lambda_k^\mathrm{NNN}-\lambda_k}.
\end{equation} 
This definition is illustrated in Fig.~\ref{fig:Poisson_vs_GinUE}-$(a)$. We then seek the probability distribution function $\varrho^{(N)}(z)$ of finding a spacing ratio with value $z$, which is defined either in the limit $N\to\infty$, or, for a finite $N$, upon averaging over spectra of an ensemble of random matrices. 

If the spectrum is real, $z\equiv r$ satisfies $-1\leq r\leq1$ and may not coincide with the ratio of consecutive spacings. 
If the spectrum is complex, $z\equiv re^{i\theta}\equiv x+i y$, with $0\leq r\leq1$, and the distribution is not necessarily isotropic. We also consider the radial and angular marginal distributions, $\varrho(r)=\int \d \theta\,r\varrho(r,\theta)$ and $\varrho(\theta)=\int \d r\,r\varrho(r,\theta)$, respectively.

We start by considering two paradigmatic cases: synthetic uncorrelated levels (corresponding to random diagonal matrices) and the Ginibre Ensembles. By natural extensions of the Berry-Tabor and Bohigas-Giannoni-Schmit conjectures, one expects integrable systems to have the same ratio statistics as uncorrelated levels and chaotic systems to follow Ginibre statistics. Because of the independence of levels in the synthetic spectrum, the presence of a reference level does not influence its two nearest neighbors and hence all ratios $z$ have the same probability, which yields a flat distribution. In contrast, for random matrices, we expect the usual repulsion, with two immediate consequences. First, the ratio density should vanish at the origin; second, the repulsion should spread all the neighbors of the reference level evenly around it, leading to a suppression of the ratio density for small angles.

Figure~\ref{fig:Poisson_vs_GinUE} shows the ratio density $\varrho(z)$ in the complex plane for uncorrelated levels, $(b)$, and GinUE matrices, $(c)$, and confirms the expectations above. For uncorrelated levels the ratio is indeed flat inside the unit circle, i.e.  $\varrho_\mathrm{Poi}(z)=(1/\pi)\Theta(1-|z|)$, with $\Theta$ the Heaviside step-function. It immediately follows that the radial and angular marginal distributions are, respectively, $\varrho_\mathrm{Poi}(\theta)=1/(2\pi)$ and $\varrho_\mathrm{Poi}(r)=2r$, and thus $\av{\cos\theta}=\int\d \theta\,\cos\theta\varrho_\mathrm{Poi}(\theta)= 0$. GinUE random matrices, on the contrary, have cubic level repulsion, $\varrho_\mathrm{GinUE}(r)\propto r^3$ as $r\to0$ (note that one power of $r$ comes from the area element on the plane), and the distribution shows some anisotropy, measured, for instance, by $\av{\cos\theta}=\int\d \theta\,\cos\theta\varrho_\mathrm{GinUE}(\theta)\simeq 0.24$.

\begin{figure}[tbp]
    \centering
    \includegraphics[width=0.99\columnwidth]{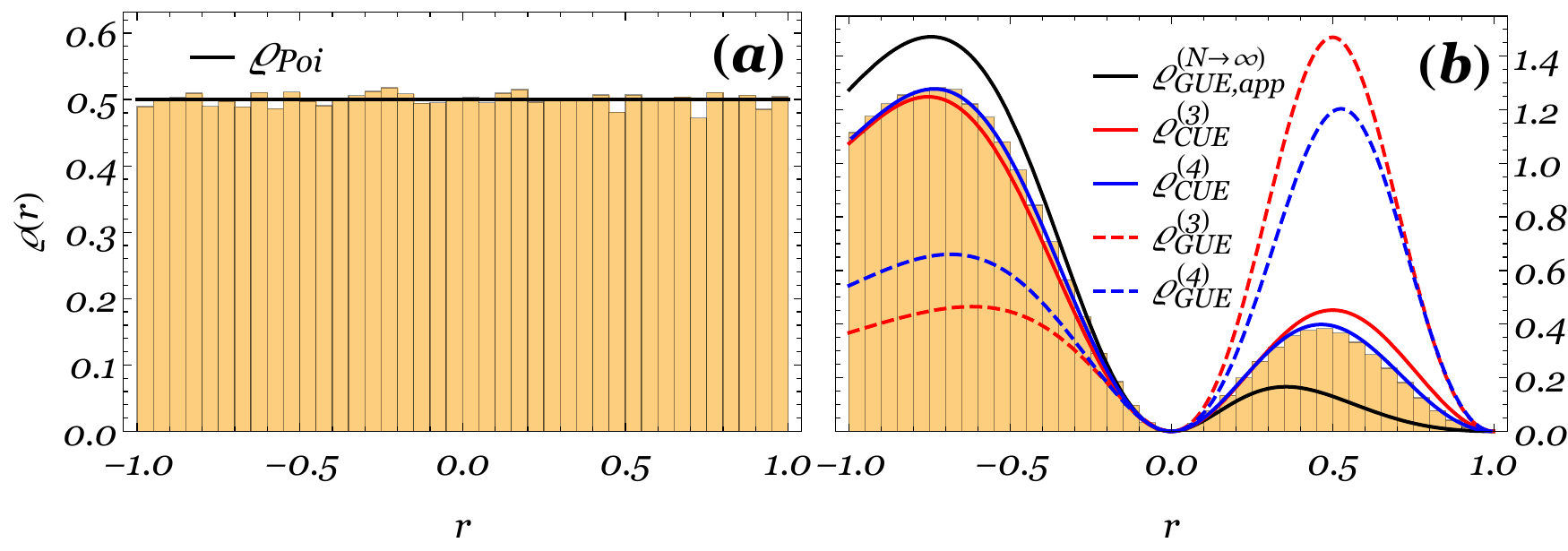}
    \caption{Comparison of numerical results and analytic predictions for the case of a real spectrum. $(a)$: Yellow bars--histogram of the ratios for $10^5$ independent levels. Black line--exact result. $(b)$: Yellow bars--histogram of the ratios obtained by exact diagonalization for $N=10^4$ GUE matrices.  Black line--approximate GUE result for $N\to\infty$, given by Eq.~(\ref{eq:gaussian_ratio_N_infty_approx}), which is valid near $r=0$. Red solid (dashed) line--exact result for $N=3$ CUE (GUE) given by Eq.~(\ref{eq:CUE_N3}) (Eq.~(\ref{eq:gaussian_ratio_N3})). Blue solid (dashed) line--exact result for $N=4$ CUE (GUE) given by Eq.~(\ref{eq:CUE_N4}) (Eq.~(\ref{eq:GUE_N4})). CUE with $N=3,4$ yield good Wigner-like surmises.}
    \label{fig:analytics_GUE}
\end{figure}

\begin{figure*}[tbp]
    \centering
    \includegraphics[width=0.99\textwidth]{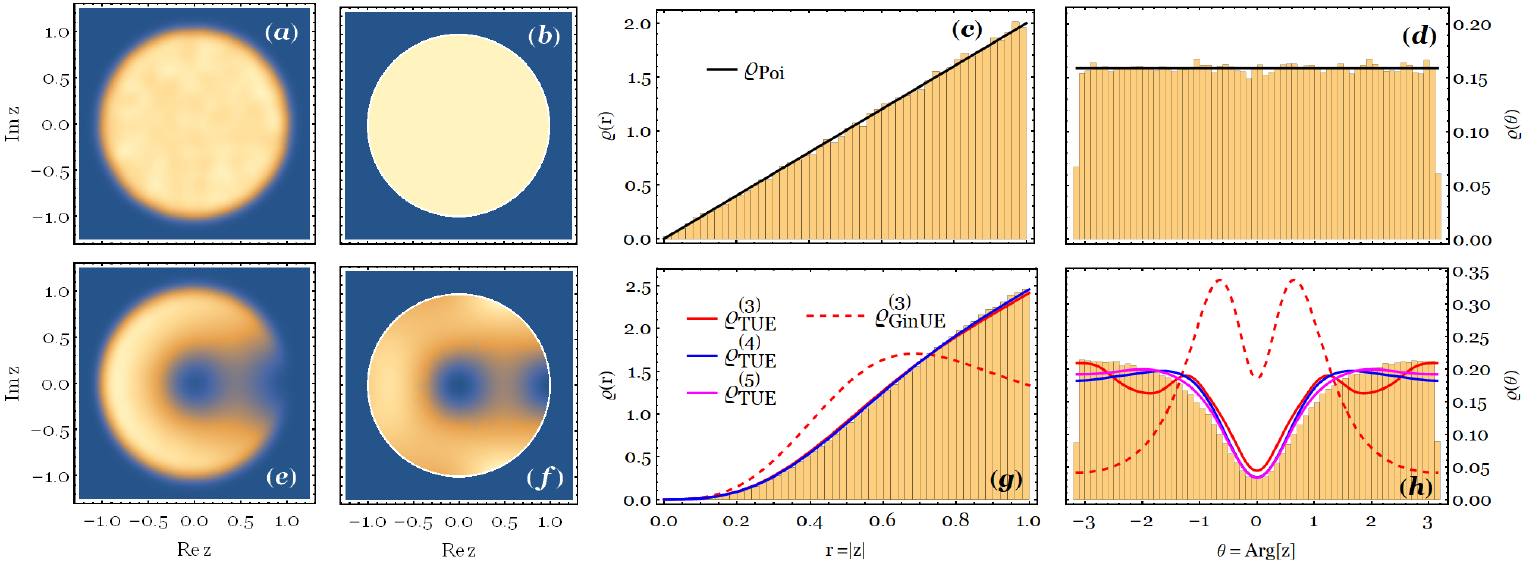}
    \caption{Distribution of complex level-spacing ratios---numerical results and analytic predictions for independent levels $(a)$--$(d)$ and GinUE-drawn matrices, $(e)$--$(h)$. $(a)$: spacing ratio density for $10^5$ independently drawn levels; $(b)$: flat distribution, Eq.~(\ref{eq:NN_NNN_ratio_r_ddim}); $(c)$ and $(d)$: histogram of $\abs{z}$ and $\arg z$ (yellow bars) and theoretical prediction (black lines); 
    $(e)$ spacing ratio distribution for GinUE matrices ($N=10^4$) obtained by exact diagonalization (ED); 
    $(f)$: surmise for the TUE with $N=3$, Eq.~(\ref{eq:TUE_N3}); 
    $(g)$ and $(h)$: histograms of $\abs{z}$ and $\arg{z}$ obtained by ED (yellow bars). Red, blue and magenta (solid) lines computed from Eq.~(\ref{eq:TUE_finite_N}) for $N=3$, $4$, $5$, respectively; dashed lines give the exact $N=3$ result from the GinUE, Eq.~(\ref{eq:GinUE_N3}), for comparison.}
    \label{fig:analytics_GinUE}
\end{figure*}

For a real (complex) spectrum, Fig.~\ref{fig:analytics_GUE}  (Fig.~\ref{fig:analytics_GinUE}) shows the distribution function of the level-spacing ratio, $z$, both for uncorrelated levels and for GUE (GinUE) random matrices of different sizes as well as the radial (radial and angular) marginal distributions. 
Contrary to the case of consecutive spacings ratios, the distribution function for small-size GUE or GinUE matrices, say with $N=3$ or $N=4$, \emph{does not} qualitatively capture the large-$N$ asymptotics, see Figs.~\ref{fig:analytics_GUE} and \ref{fig:analytics_GinUE}, respectively. For a complex-valued spectrum, Figs.~\ref{fig:analytics_GinUE}-$(g)$ and $(h)$ show that in the GinUE distribution for small $N$ (dashed red line) there is an enhancement of the small angles, rather than the suppression seen at large $N$ (yellow histogram). A similar issue arises for the case of a real spectrum shown in Fig.~\ref{fig:analytics_GUE}-$(b)$: for large $N$ (yellow histogram), there is a high probability of finding negative ratios, while for small $N$ (red and blue dashed lines), the probability of positive $r$ is higher. 

This small-$N$ \emph{peak inversion} can be understood as a boundary effect. For definiteness, consider matrices drawn from a Hermitian ensemble. For $N=3$, the sign of the ratios is completely fixed: The two levels at the edges must, by construction, have both neighbors on the same side and hence $r > 0$; the central level has one neighbor on each side and hence $r < 0$; it follows that the area below the negative-$r$ peak is $1/3$ and the area below the positive-$r$ peak is $2/3$ (the analytical expressions below confirm this reasoning exactly) As $N$ increases, the edge levels, which always have positive ratios, looe importance relative to the growing number of bulk levels, which tend to have negative ratios, and peak inversion follows. The argument for non-Hermitian matrices is analogous: bulk levels favor large angles while boundary levels lead to small angles; at small $N$, boundary levels dominate, but they cannot compete in number with bulk levels at large $N$. 

\begin{figure}
    \centering
    \includegraphics{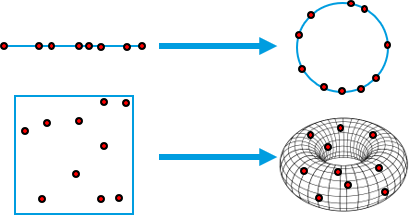}
    \caption{Sketch of how to eliminate boundary effects that preclude small-$N$ surmises of complex spacing ratio statistics. (a) Instead of computing the ratios of the GUE, we compute those of the circular unitary ensemble; (b) instead of computing the ratios of the GinUE, we compute those of the toric unitary ensemble (note that this representation is only schematic, as the two-torus is embedded in $\mathbb{R}^4$).}
    \label{fig:periodic_boundary}
\end{figure}

The strong $N$-dependence thus precludes any small-size Wigner-like surmise using GinUE-drawn matrices.
One of our main results is that these boundary effects can be overcome by using different ensembles with the same asymptotic large-$N$ distribution. 
Figure~\ref{fig:periodic_boundary} sketches the main idea of our approach.
For a real spectrum, we obtain a surmise using the spacing ratios of the circular unitary ensemble (CUE)~\cite{dyson1962i,haake2013}, whose spectrum lies along the unit circle, therefore avoiding boundary effects.
Figure~\ref{fig:analytics_GUE}-$(b)$ shows that the predictions of this method (solid red and blue lines) converge rapidly for increasing $N$ and already give a very good quantitative agreement for $N=3$ and $N=4$. 
The toric unitary ensemble (TUE), introduced in the next section, generalizes this idea for the case of a complex spectrum.  
Figs.~\ref{fig:analytics_GinUE}-$(g)$ and $(h)$ show that the predictions obtained in this way for small $N$ (solid and red lines) also qualitatively reproduce the large-$N$ results.

A second main result of our work is to verify that these distributions do generalize the Berry-Tabor and Bohigas-Giannoni-Schmit conjectures to physical situations where the relevant operators have complex-valued spectra.
By studying different physical models where non-Hermitian matrices arise, we show that known integrable cases have a flat distribution of complex spacing ratios whereas generic cases, which are expected to be chaotic, conform to Random Matrix Theory predictions. 
Figures~\ref{fig:Liouvillian_ratio}-$(a)$--$(e)$, below, illustrate our findings for a spin-$1/2$ chain, subject to boundary driving and/or bulk dissipation, modeled by Markovian Lindblad dynamics. The flat distribution of Fig.~\ref{fig:Liouvillian_ratio}-$(a)$, corresponds to a boundary driven XX chain with bulk dephasing, which is known to have an integrable Liouvilian. This case contrasts with the nonintegrable cases, $(b)$--$(e)$, where the distribution of complex spacing ratios is highly asymmetric and is expected to reach the GinUE distribution in the thermodynamic limit. Similar results are reported in Sec.~\ref{subsection:examples_MBL} for the case of nonunitary Hamiltonian dynamics, and in Sec.~\ref{subsection:examples_classical_Markov} for the spectrum of the Markov matrix describing the ASEP. 
These results provide solid evidence that the complex level-spacing ratio distribution can be used to distinguish chaotic from integrable dynamics of operators with complex-valued spectra.  

\section{Analytical results: exact distribution functions and surmises}
\label{sec:Surmises}

In this section, we summarize our main analytical results regarding the complex spacing ratio distribution of independent levels and the small-$N$ surmises obtained for the CUE and the TUE. 

For independent levels the spacing ratios are isotropic. Therefore, the only nontrivial distribution is that of $r=\abs{z}$, which can be obtained analytically for $d$ dimensions (generalizing real, $d=1$, and complex, $d=2$, spectra). Furthermore, all joint-spacing distributions of more than one spacing factorize into single-spacing distributions $\hat{P}(s)$ and one can write the ratio distribution in terms of $\hat{P}(s)$ only:
\begin{equation}\label{eq:P(r)_definition}
    \varrho_\mathrm{Poi}(r)=
    \Theta(1-r)\int_0^\infty \d s\, \frac{s\,\hat{P}(s)\,\hat{P}(rs)}{\int_{rs}^\infty\d s'\, \hat{P}(s')}\,.
\end{equation}
At the unfolded scale, the $d$-dimensional single-spacing distribution $\hat{P}(s)$ is a Brody distribution~\cite{brody1973},
\begin{equation}
    \hat{P}(s)=d\,\Gamma(1+1/d)^ds^{d-1}e^{-\Gamma(1+1/d)^d\,s^d}\,,
\end{equation}
which recovers the standard exponential distribution for one-dimensional spectra. The ratio distribution in $d$ dimensions,
\begin{equation}\label{eq:NN_NNN_ratio_r_ddim}
    \varrho_\mathrm{Poi}(r)=d\, r^{d-1}\heav{1-r},
\end{equation}
then follows. This shows that (after introducing a $d$-dimensional volume element) the ratio distribution is, indeed, flat. For more details on spacing ratios for uncorrelated random variables, and some generalizations, see Appendix~\ref{sec:poisson}.

We now address random matrix ensembles starting with the case of real spectra. 
The level-spacing ratio distribution function for $N\times N$ matrices drawn from arbitrary Hermitian ensembles, $\varrho^{(N)}(r)$, can be formally written as an $(N-1)$-fold integral over the joint eigenvalue distribution function [Eq.~(\ref{eq:hermitian_ratio_general})].
By specializing to the Gaussian ensembles, this quantity can be explicitly computed for small-size matrices, e.g.\ $N=3$ [Eq.~(\ref{eq:gaussian_ratio_N3})]. Other small sizes are still amenable to a brute-force evaluation of the integrals. However, we were not able to determine the complete asymptotic large-$N$ distribution using this approach. 
Nonetheless, it can be employed to capture the scaling,  $\varrho^{(N\to\infty)}_\mathrm{GUE}(r) \propto r^\beta $, in the vicinity of $r=0$. 

As shown in the last section, although larger values of $N$ suppress the weight of boundary effects, the convergence towards the infinite-$N$ limit is very slow. 
Convergence is much faster in the case of the circular ensembles (CE), where results for small-size matrices ($N=3$, $N=4$) from the CE already capture most of the features of the large-$N$ asymptotics. Since for $N\to\infty$, CE and GE have the same level-spacing ratio statistics, we can use CE small-size matrices as surmises for the GE large-$N$ distribution.
As for GE, for CE the level-spacing ratio distribution function for $N\times N$ matrices can be formally obtained in the form of an $(N-1)$-fold integral. 
For $N=3$, the ratio distribution reads
\begin{equation}\label{eq:CUE_N3_integral}
\begin{split}
    \varrho^{(3)}_\mathrm{CUE}(r)&\propto\heav{1-r^2}\int_{-\pi}^\pi\d v \abs{v}\left(1-\cos v\right)\\
    &\times\left(1-\cos rv\right)\left(1-\cos(r-1)v\right),
\end{split}
\end{equation}
which is evaluated in Eq.~(\ref{eq:CUE_N3}), yielding a ratio of polynomials of $r$, whose explicit form is given in the Supplemental Material~\cite{SM}. 
A similar expression was also obtained for $N=4$ [Eq.~(\ref{eq:CUE_N4})]. For further details on Hermitian ensembles, we refer the reader to Appendix~\ref{sec:analytics_real}.

Finally, we turn to non-Hermitian ensembles, considering, for simplicity, only the case $\beta=2$.  The general expression of the ratio distribution, for an arbitrary ensemble, is a $2(N-1)$-fold real integral over the ensemble's joint eigenvalue distribution [Eq.~(\ref{eq:nonhermitian_ratio_general})]. For the GinUE, the distribution for $N=3$ can be computed explicitly [Eq.~(\ref{eq:GinUE_N3})], but, again, it does not correctly describe the large-$N$ asymptotics. The leading-order expansion in powers of $r$ yields $\varrho^{(N)}(r) \propto r^{3}$, but is valid only around $r=0$.

In order to eliminate boundary effects from a complex spectrum we consider the two-dimensional analogue of the circular ensemble. This novel ensemble has eigenvalues equally distributed on the two-dimensional (Clifford) torus, $\mathbb{T}^2=\mathbb{S}^1\times\mathbb{S}^1\subset \mathbb{S}^3\subset\mathbb{R}^4$, which can be parametrized by two angles, $\vartheta\in(-\pi,\pi]$, $\varphi\in(-\pi,\pi]$. In analogy with the CUE, we dubbed it the Toric Unitary Ensemble (TUE). Therefore, $P^{(N)}_\mathrm{TUE}$, is flat on the torus. It follows that $P^{(N)}_\mathrm{TUE}$ is fully determined by the Vandermonde interaction on the torus, and it reads
\begin{equation}\label{eq:TUE_joint}
\begin{split}
    P^{(N)}_\mathrm{TUE}&(\vartheta_1,\dots,\vartheta_N;\varphi_1,\dots,\varphi_N)\\
    &\propto \prod_{j<k}\left[2-\cos(\vartheta_j-\vartheta_k)-\cos(\varphi_j-\varphi_k)\right].
\end{split}
\end{equation}
Setting $N=3$, we compute a Wigner-like surmise for the complex spacing ratio distribution for non-Hermitian random matrices,
\begin{equation}\label{eq:TUE_N3}
\begin{split}
    &\varrho^{(3)}_\mathrm{TUE}(x,y)\propto\int_{-\pi}^\pi\d s\d t (s^2+t^2)^2\left[2-\cos s-\cos t\right]\\
    &\times\left[2-\cos(sx-ty)-\cos(t x+s y)\right]\\
    &\times\left[2-\cos(s(x-1)-ty)-\cos(t (x-1)+s y)\right].
\end{split}
\end{equation}
The integral of Eq.~(\ref{eq:TUE_N3}) and its generalizations for $N=4,5,\dots$ [see  Eq.~(\ref{eq:TUE_finite_N})], can be numerically integrated and provide our surmises for the large-$N$ asymptotics of the GinUE universality class. Figs.~\ref{fig:analytics_GinUE}-$(e)$--$(h)$ show that the convergence of the radial marginal distribution is similar to that of the real case: both $N=3$ and $N=4$ provide good approximations, the latter being already almost indistinguishable from large-$N$ exact diagonalization data. The angular marginal distribution has a much slower convergence, especially near $\theta=\pm\pi$. Although the qualitative features are already captured for $N=3$, quantitatively, one can still distinguish the discrepancies even for $N=5$ in Fig.~\ref{fig:analytics_GinUE}-$(h)$, although the agreement does improve as $N$ increases. For further details on non-Hermitian ensembles, see Appendix~\ref{sec:analytics_complex}.

\begin{figure*}[tbp]
    \centering
    \includegraphics[width=0.75\textwidth]{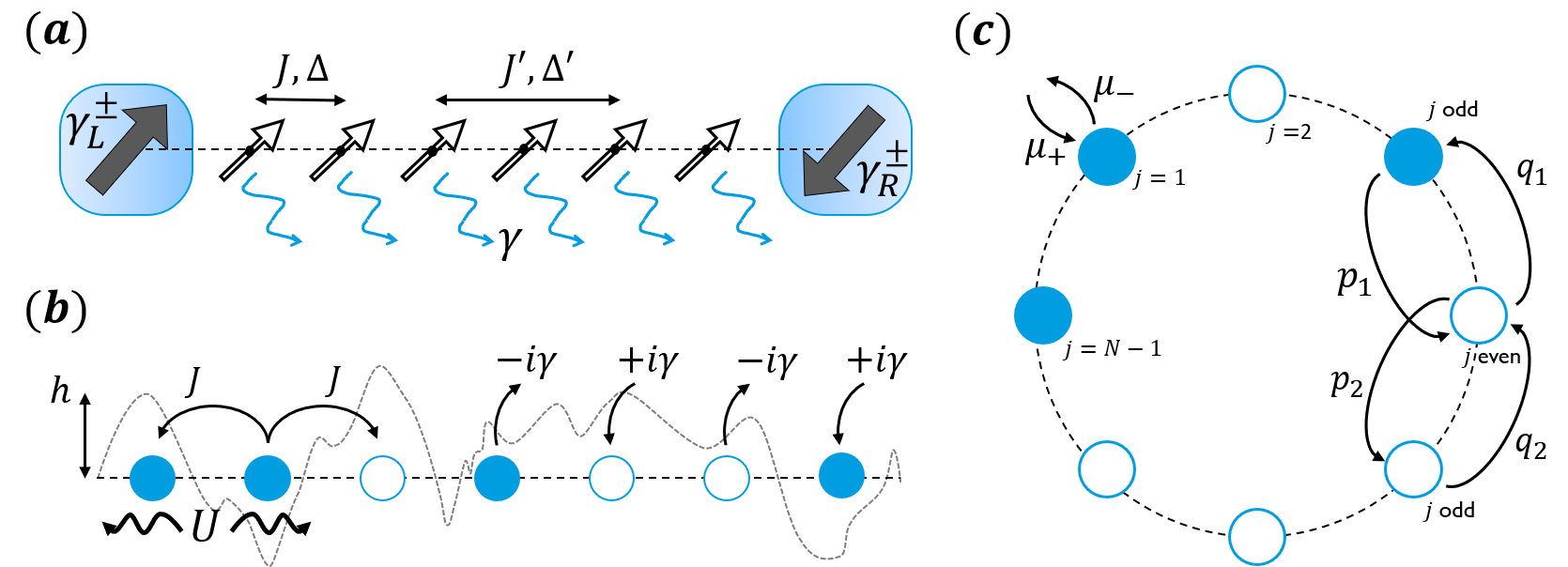}
    \caption{Sketch of the three models studied: $(a)$ a boundary-driven dissipative spin chain, Sec.~\ref{subsection:examples_spin_chains}; $(b)$ a non-Hermitian disordered many-body system, Sec.~\ref{subsection:examples_MBL}; $(c)$ a classical simple exclusion process, Sec.~\ref{subsection:examples_classical_Markov}.}
    \label{fig:sketches_examples}
\end{figure*}

\section{Physical Applications}
\label{sec:examples}

We now determine the complex spacing ratio distribution of several different numerical examples of current interest. In Sec.~\ref{subsection:examples_spin_chains} we consider the Lindbladian description of boundary-driven dissipative spin-chains, in Sec.~\ref{subsection:examples_MBL} we address a non-Hermitian Hamiltonian modeling many-body localization, and in Sec.~\ref{subsection:examples_classical_Markov} we study a classical stochastic process.

\subsection{Boundary-driven dissipative spin-chains}
\label{subsection:examples_spin_chains}

A simple way of modeling open quantum systems is by employing a master equation approach to describe the dynamics of the system's reduced density matrix. When the environment is Markovian, this procedure substantially simplifies and the master equation acquires the Lindblad form
\begin{equation}
\begin{split}
    &\frac{\d}{\d t}\rho(t)=\mathcal{L}\rho(t)\\
    &\equiv-i\,\comm{H}{\rho(t)}+\sum_{\mu=1}^D\left(W_\mu\rho(t)W_\mu^\dagger-\frac{1}{2}\acomm{W_\mu^\dagger W_\mu}{\rho(t)}\right),
\end{split} \label{eq:Lindblad}
\end{equation}
where $H$ is the Hamiltonian and $W_\mu$, with $\mu=1,\dots,D$, are called jump operators, modeling the system-environment interaction. 

Here, we study the spectrum of a family of non-Hermitian operators $\mathcal{L}$ for a well-studied physical setup of a chain of spins-$1/2$. In the middle of the chain, the magnetization along $z$ is conserved and the net role of the environment is to dephase the system, i.e. decrease off-diagonal amplitudes of the density matrix when written in the $z$-basis. 
At the two ends of the chain, the spin magnetization can be injected or extracted at fixed rates. 
This model had been extensively used for studying nonequilibrium spin transport~\cite{prosen2011,buca2012,medvedyeva2016}. 

\subsubsection{Model}
We consider a chain of $N$ spins $1/2$ evolving in time by the action of a Lindblad-Liouvillian operator, given by Eq.~(\ref{eq:Lindblad}), and schematically represented in Fig.~\ref{fig:sketches_examples}-$(a)$. $H$ belongs to a family of next-to-nearest-neighbor Heisenberg XXZ Hamiltonians,
\begin{equation}
\begin{split}
    H&=J\sum_{\ell=1}^{N-1}\left(\sigma_\ell^\mathrm{x}\sigma_{\ell+1}^\mathrm{x}+\sigma_\ell^\mathrm{y}\sigma_{\ell+1}^\mathrm{y}+\Delta\sigma_\ell^\mathrm{z}\sigma_{\ell+1}^\mathrm{z}\right)\\
    &+J'\sum_{\ell=1}^{N-2}\left(\sigma_\ell^\mathrm{x}\sigma_{\ell+2}^\mathrm{x}+\sigma_\ell^\mathrm{y}\sigma_{\ell+2}^\mathrm{y}+\Delta'\sigma_\ell^\mathrm{z}\sigma_{\ell+2}^\mathrm{z}\right),
\end{split}
\end{equation}
with $\sigma_\ell^\alpha$ the Pauli operators, $\alpha\in\{\mathrm{x},\mathrm{y},\mathrm{z}\}$ and $\ell\in\{1, 2\dots,N\}$, and
$J$ ($J'$)  the nearest- (next-to-nearest-) neighbor exchange coupling and $z$-axis anisotropy $\Delta$ ($\Delta'$).  
To model bulk dephasing and spin injection, we consider two types of incoherent jump processes (in total, $D=N+4$ of them):
\begin{enumerate}[(i)]
    \item bulk dephasing of all spins,
    \begin{equation}
        W_\ell=\sqrt{\gamma}\sigma_\ell^\mathrm{z},\quad\ell\in\{1,\dots,N\}\,;
    \end{equation}
    \item amplitude damping (spin polarization) processes at the boundaries,
    \begin{equation}
    \begin{split}
        &W_{N+1}=\sqrt{\gamma_\mathrm{L}^+}\sigma_1^+\,,\quad 
        W_{N+2}=\sqrt{\gamma_\mathrm{L}^-}\sigma_1^-\,,\\ 
        &W_{N+3}=\sqrt{\gamma_\mathrm{R}^+}\sigma_N^+\,,\quad
        W_{N+4}=\sqrt{\gamma_\mathrm{R}^-}\sigma_N^-\,.
    \end{split}
    \end{equation}
\end{enumerate}
Here, $\gamma$ controls the dephasing rate and $\gamma_{\text{L}/\text{R}}^{\pm}$ controls the spin injection ($+$) and extraction ($-$) at the left ($\text{L}$) or right ($\text{R}$) ends of the chain.  
Thus, the model is characterized by the nine parameters $J$, $J'$, $\Delta$, $\Delta'$, $\gamma$, $\gamma_\mathrm{L,R}^\pm$, which allows us to tune its integrability or chaoticity.

\begin{figure*}[tbp]
    \centering
    \includegraphics[width=0.99\textwidth]{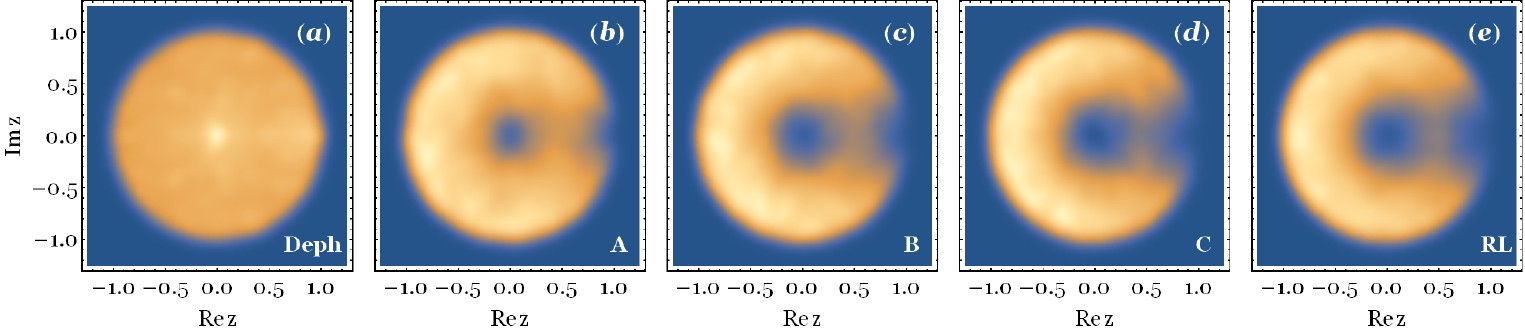}
    \caption{Complex spacing ratio density for different Liouvillian spectra. $(a)$ Deph--boundary driven XX chain with bulk dephasing; $(b)$ A--XXX chain with pure-source/pure-sink driving; $(c)$ B--XXX chain with arbitrary polarizing boundary driving; $(d)$ C--XXZ chain with nearest neighbor and next-to-nearest-neighbor interactions; $(e)$ RL--random Liouvillian~\cite{sa2019} at strong dissipation. The spin-chain Liouvillians were diagonalized for $N=10$, $M=7$.}
    \label{fig:Liouvillian_ratio}
\end{figure*}

The Hilbert space is spanned by the states $\ket{s_1,\dots,s_N}$, with $s_\ell= \pm 1 $. The space of density matrices--the Liouville space, $\mathcal{K}$--in which $\mathcal{L}$ acts, is spanned by $\kket{s_1,\dots,s_N;s'_1,\dots,s'_N}=\ket{s_1,\dots,s_N}\otimes\bra{s'_1,\dots,s'_N}^{\sf{T}}$. 
Using this notation, we formulate the spectral problem for the Liouvillian superoperator $\mathcal{L}$ in terms of a $4^N\times 4^N$ matrix representation acting on a $4^N$-dimensional density operators $\rho\in\mathcal{K}$,
\begin{equation}\label{eq:spin_chain_Liouvillian}
\begin{split}
    &\mathcal{L}=-i\Bigg\{\left(H-\frac{i}{2}\sum_{\mu=1}^rW_\mu^\dagger W_\mu\right)\otimes\mathbbm{1}\\
    &-\mathbbm{1}\otimes\left(H+\frac{i}{2}\sum_{\mu=1}^rW_\mu^\dagger W_\mu\right)^{\sf{T}}\Bigg\}+\sum_{\mu=1}^DW_\mu\otimes W_\mu^*\,.
\end{split}
\end{equation}

The superoperator $\mathcal{S}^\mathrm{z}=S^\mathrm{z}\otimes \mathbbm{1}-\mathbbm{1}\otimes S^{\mathrm{z} \sf{T}}$, with $S^\mathrm{z}=\sum_{\ell=1}^N\sigma_\ell^\mathrm{z}$ the total z-axis magnetization, commutes with the Liouvillian $\comm{\mathcal{L}}{\mathcal{S}^{\mathrm{z}}}=0$~\cite{buca2012}.
This result implies that $\mathcal{K}$ splits into sectors, $\mathcal{K}_M$, of conserved quantum number $M$, each spanned by $\binom{2N}{M}$ states $\kket{s_1,\dots,s_N;s'_1,\dots,s'_N}$ with $\sum_\ell(s_\ell-s'_\ell)=N-M$. The tensor-product representation of the Liouvillian block-diagonalizes into $2N+1$ sectors $\mathcal{L}_M$, $\mathcal{L}=\bigoplus_{M=0}^{2N} \mathcal{L}_M$, with each block a $\binom{2N}{M}\times\binom{2N}{M}$ matrix. The symmetric sector $M=N$ contains all states with vanishing magnetization, including the steady state. 

Note that, for $M \neq N$, each complex conjugate pair of eigenvalues of the Liouvillian is divided across two sectors of symmetric magnetization, i.e.\ if sector $M$ contains the eigenvalue $\Lambda$, then sector $\abs{2N-M}$ contains the eigenvalue $\Lambda^*$. The different sectors $M$ must be analyzed separately because spectra corresponding to different conserved quantum numbers form independent level sequences that superimpose without interacting~\cite{guhr1998}.

\subsubsection{Numerical results}

\begin{table*}[htbp]
    \centering
    \caption{Single-number signatures of integrability/chaos for different Liouvillians: models Deph, A, B, C, and RL. They are compared with exact analytical results for uncorrelated random variables (labeled Poisson), numerical exact diagonalization of ($10^4\times10^4$) random GinUE matrices, and TUE surmise estimates for $N=3,4,5$ (subscripts denote matrix size) computed from Eq.~(\ref{eq:TUE_finite_N}). The convergence of $\av{\cos\theta}$ computed from the TUE surmises is much slower than that of $\av{r}$, as noted in the text.}
    \label{tab:Liouvillian_signatures}
    \vspace{+1ex}
    \begin{tabular}{cclllllllll}
        \hline
        & Poisson & 
        \multicolumn{1}{c}{Deph} & 
        \multicolumn{1}{c}{A} & 
        \multicolumn{1}{c}{B} & 
        \multicolumn{1}{c}{C} & 
        \multicolumn{1}{c}{RL} & 
        \multicolumn{1}{c}{GinUE$_{10^4}$}  & \multicolumn{1}{c}{TUE$_3$}  &
        \multicolumn{1}{c}{TUE$_4$}  &
        \multicolumn{1}{c}{TUE$_5$}
        \\
        \hline
        $-\langle\cos\theta\rangle$ &
        $0$ & $-0.0305(26)$ & $0.1293(24)$ & $0.1890(23)$ & $0.2349(7)$ & $0.2287(20)$ & $0.24051(61)$ & $0.15322(1)$ & $0.1695(4)$ & $0.1938(86)$\\
        $\langle r \rangle$ &
        $2/3$ & \ \ \,$0.6537(9)$ & $0.7122(7)$ & $0.7292(7)$ & $0.7368(7)$ & $0.7373(6)$ & $0.73810(18)$ & $0.73193(1)$ & $0.73491(5)$ & $0.7315(50)$\\
        \hline
    \end{tabular}
\end{table*}

Numerical results were obtained by exactly diagonalizing the matrix representation of $\mathcal{L}$, Eq.~(\ref{eq:spin_chain_Liouvillian}), for different chain length $N$ and in specific sectors $M$. The largest system we diagonalized was $N=10$ spins in the sector with magnetization $M=7$, which corresponds to a $77520\times 77520$ matrix.
The following four cases of parameters were studied:
\begin{itemize}
    \item \textbf{(Deph)} Boundary driven XX chain with bulk dephasing. Numerical parameters chosen as $J=1$, $J'=\Delta=\Delta'=0$, $\gamma=1$, $\gamma_\mathrm{L}^+=0.5$, $\gamma_\mathrm{L}^-=1.2$, $\gamma_\mathrm{R}^+=1$, $\gamma_\mathrm{R}^-=0.8$. This model can be mapped onto the Fermi-Hubbard model with imaginary interaction $U=i\gamma$~\cite{medvedyeva2016} and hence is Bethe-ansatz integrable.
    \item \textbf{(A)} Isotropic Heisenberg (XXX) chain  with pure-source/pure-sink driving and no dephasing. Numerical parameters chosen as $J=\Delta=1$, $J'=\Delta'=0$, $\gamma=\gamma_\mathrm{L}^-=\gamma_\mathrm{R}^+=0$, $\gamma_\mathrm{L}^+=0.6$, $\gamma_\mathrm{R}^-=1.4$. The steady state of this model is known to be integrable~\cite{prosen2011}, but the bulk of the spectrum is likely not integrable.
    \item \textbf{(B)} XXX chain  with arbitrary boundary-driving and no dephasing. Numerical parameters chosen as $J=\Delta=1$, $J'=\Delta'=0$, $\gamma=0$, $\gamma_\mathrm{L}^+=0.5$, $\gamma_\mathrm{L}^-=0.3$ $\gamma_\mathrm{R}^+=0.3$, $\gamma_\mathrm{R}^-=0.9$. The bulk Hamiltonian of this model is integrable, but, by adding a generic boundary-driving, not even the steady state is expected to be exactly-solvable.
    \item \textbf{(C)} XXZ chain with next-to-nearest-neighbor interactions, arbitrary boundary-driving, and no dephasing. Numerical parameters chosen as $J=J'=1$, $\Delta=0.5$, $\Delta'=1.5$, $\gamma=0$, $\gamma_\mathrm{L}^+=0.5$, $\gamma_\mathrm{L}^-=0.3$ $\gamma_\mathrm{R}^+=0.3$, $\gamma_\mathrm{R}^-=0.9$. For this model, not even the bulk Hamiltonian is integrable.
\end{itemize}
Additionally, we considered a fifth model for comparison:
\begin{itemize}
    \item \textbf{(RL)} Random Liouvillian~\cite{sa2019} at strong dissipation. Numerical parameters (adopting the notation of Ref.~\cite{sa2019}) chosen as $N=80$, $\beta=2$, $r=2$, $g=100$.
\end{itemize}

We applied the procedure described at the beginning of Sec.~\ref{sec:main_results} to compute the distribution of the complex spacing ratios for the five models depicted in Fig.~\ref{fig:Liouvillian_ratio}. There is a striking difference between the integrable model (Deph), and the others, which are expected to be chaotic.  
The dephasing-XX model, Fig.~\ref{fig:Liouvillian_ratio}-$(a)$, displays a distribution similar to that of uncorrelated levels. Models B, C, and RL, Figs.~\ref{fig:Liouvillian_ratio}-$(c)$, $(d)$, $(e)$, respectively, clearly conform to RMT statistics. Model A, Fig.~\ref{fig:Liouvillian_ratio}-$(b)$, on the other hand, shows an intermediate behavior between Poisson and RMT statistics, both in terms of radial level repulsion and of anisotropy of the angular distribution. This could arise either from actual intermediate statistics of the spectrum or from finite-size effects. On the contrary, model C already displays the universal large-$N$ behavior, with no noticeable finite-size effects.
These results indicate that complex spacing ratios indeed offer a clean and simple signature of quantum chaos in Markovian setups. 

\begin{figure}[tbp]
    \centering
    \includegraphics[width=0.99\columnwidth]{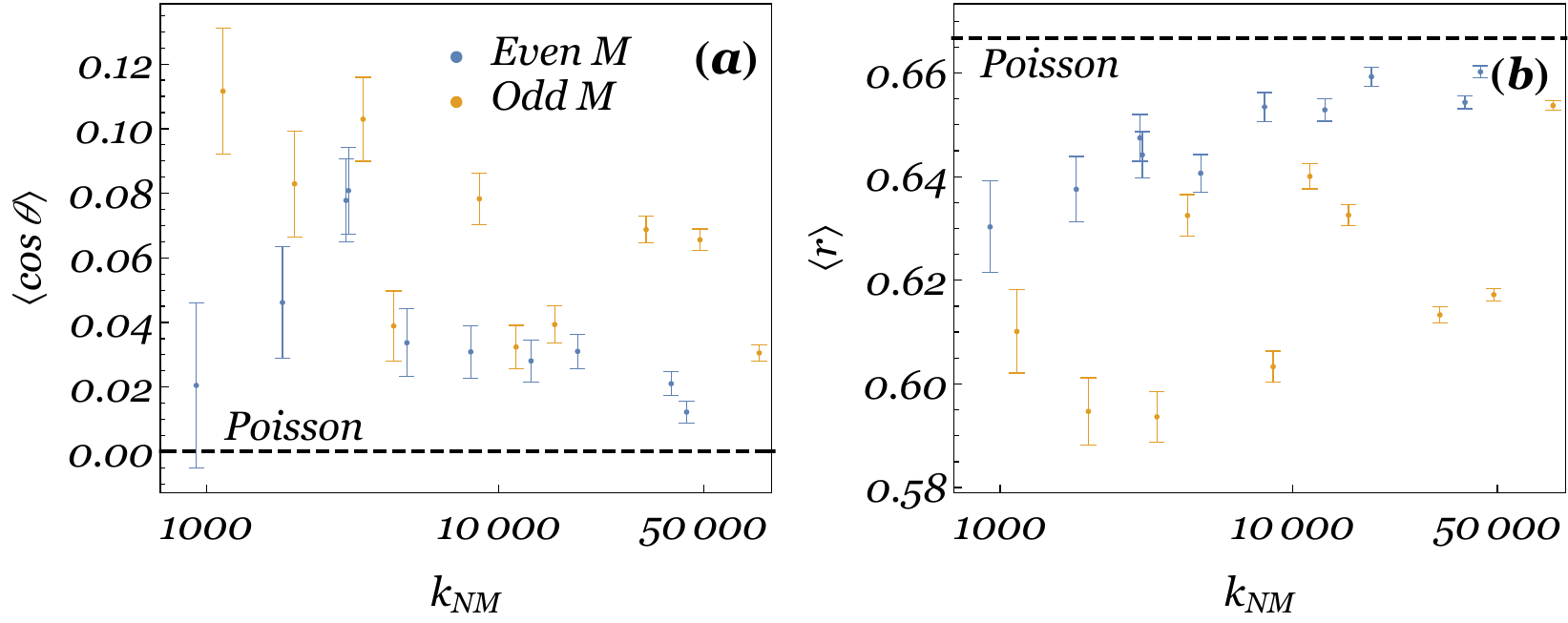}
    \caption{Finite-size effects on the complex spacing ratios of a spin-chain Liouvillian of the dephasing-XX model, for different chain lengths $N$ and spin sectors $M$, the sector dimension being $k_{N\!M}=\binom{2N}{M}$. $(a)$: average value of $\cos\theta$; $(b)$: average value of $r$. The upper (lower) dashed line corresponds to the GinUE- (Poisson-) limit.}
    \label{fig:LDeph_finite_size}
\end{figure}

\subsubsection{Single-number signatures}

Next, we try to capture the main features of the distribution of complex spacing ratios through a reduced set of numbers, which we call \emph{single-number signatures}. 
A popular single-number signature, used for the ratio of undirected spacings, is the degree of level repulsion $\alpha$, i.e.\ the exponent describing the power-law behavior of the radial marginal distribution, $\varrho(r)\propto r^\alpha$, as $r\to0$ or, equivalently, $\alpha=\lim_{r\to0}\log\varrho(r)/\log r$. For Hermitian random matrices it is given by the Dyson index, $\alpha=\beta$; for non-Hermitian random matrices from the universality class of either GinOE, GinUE, or GinSE it is $\alpha=3$; while for real independent random variables it is $\alpha=0$; and for complex uncorrelated random variables it is $\alpha=1$. Although the degrees of repulsion $\alpha$ just stated can be easily checked against the numerical spectra and the above predictions confirmed, an actual computation of $\alpha$ for a given spectrum introduces a large relative error. An alternative measure of the radial distribution is given by its moments, for instance, the mean $\av{r}$. For independent random variables, we can compute exactly $\av{r}=2/3$, while for GinUE matrices we numerically find $\av{r}\approx0.74$. To measure the anisotropy of the angular marginal distribution, we consider $\av{\cos\theta}$, which is zero for a flat distribution and positive (negative) when small angles are enhanced (suppressed), in particular, $\av{\cos\theta}\approx-0.24$ for large-$N$ GinUE matrices.

We give the values of $\av{\cos\theta}$ and $\av{r}$ for the five Liouvillians in Table~\ref{tab:Liouvillian_signatures} (the spin-chain Liouvillian values are for $N=10$, $M=7$). From the radial measure $\av{r}$ it is difficult to discern the integrability or chaoticity of the different models. Indeed, the values for all four models A, B, C, RL are within $3\%$ of each other. On the contrary, as anticipated in Sec.~\ref{sec:main_results}, the angular distribution offers a more sensitive signature. From the value of $\av{\cos\theta}$, the dephasing-XX model clearly supports Poisson statistics and models C and RL are very close to RMT statistics. Model B, which also seemed very close to RMT statistics from Fig.~\ref{fig:Liouvillian_ratio} and from the value of $\av{r}$ here shows a more significant deviation. Finally, model A has a value of $\av{\cos\theta}$ almost exactly halfway between uncorrelated levels and RMT statistics, attesting to its intermediate behavior, at least for the sector dimensions considered.

\begin{figure}[tbp]
    \centering
    \includegraphics[width=0.99\columnwidth]{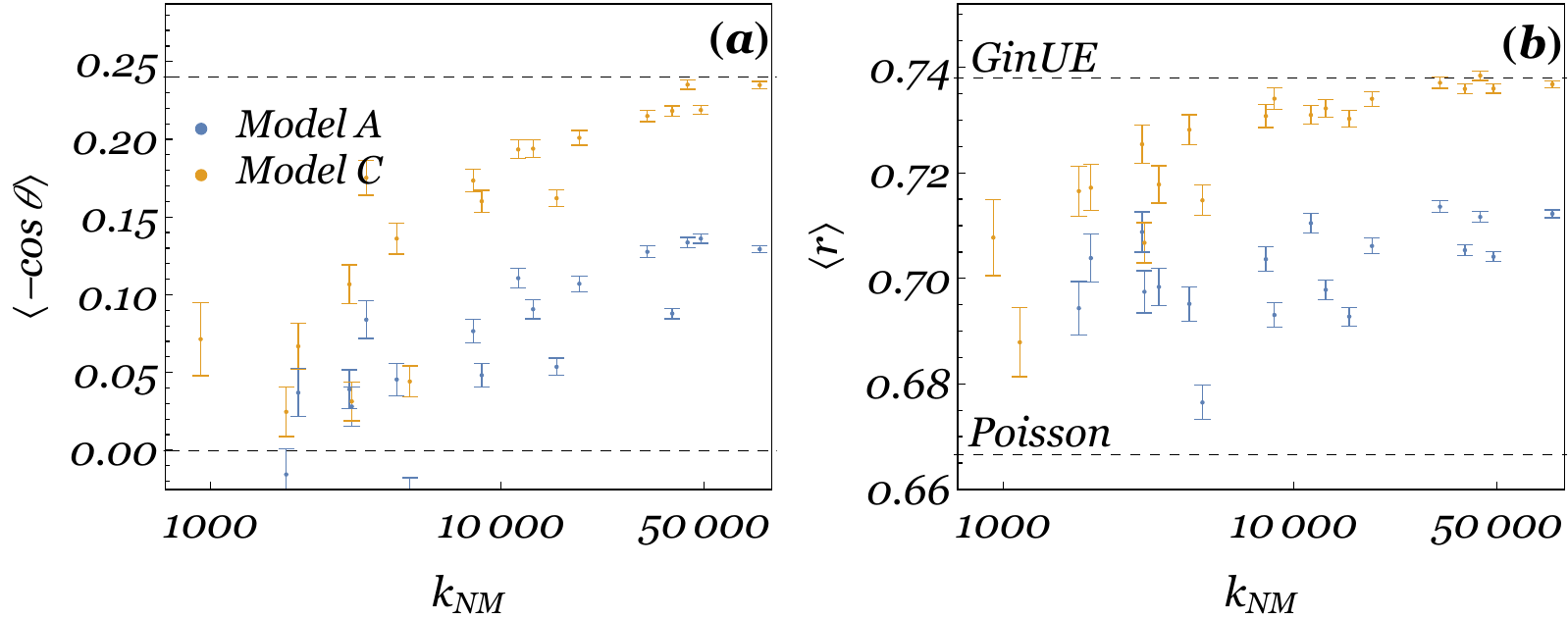}
    \caption{Finite-size effects on the complex spacing ratios of spin-chain Liouvillians of model A (blue) and model C (orange). We consider different chain lengths $N$ and spin sectors $M$, the sector dimension being $k_{N\!M}=\binom{2N}{M}$. $(a)$: average value of $\cos\theta$; $(b)$: average value of $r$. The upper (lower) dashed line corresponds to the GinUE- (Poisson-) limit.}
    \label{fig:LAC_finite_size}
\end{figure}

\subsubsection{Finite-size scaling}
We now provide a finite-size analysis of the dephasing-XX model---which conforms to Poisson level statistics---, model A---with intermediate statistics---, and model C---with RMT statistics.

\textit{Dephasing-XX model.---}
We considered single-number signatures $\av{\cos\theta}~$ and $\av{r}$ as a function of sector dimension in Fig.~\ref{fig:LDeph_finite_size}. Both signatures clearly tend to the expected value for uncorrelated random variables (dashed line) as $k_{N\!M}$ increases. There is also a visible difference between sectors with even or odd $M$, with sectors of even $M$ tending faster to the large-dimension universal limit. This aspect is also visible in Fig.~\ref{fig:LDeph_finite_size}-$(d)$.

\textit{Model A.---}
We observed that $(i)$ there is a smaller degree of level repulsion here than for fully chaotic systems (and which does not increase substantially when $k_{N\!M}$ grows by nearly two orders of magnitude) and $(ii)$ some anisotropy is developing as $k_{N\!M}$ increases. In Fig.~\ref{fig:LAC_finite_size} we plot the two single-number signatures $\av{\cos\theta}$ and $\av{r}$. While the anisotropy indeed grows (slowly) with $k_{N\!M}$, the average of the radial marginal distribution is approximately flat. No difference between even $M$ and odd $M$ is visible in this case. Contrary to the dephasing-XX model above, for which the $N=10$, $M=7$ sector is already very close to the limiting Poisson statistics, the convergence of model A towards either Poisson or GinUE statistics is much slower. 
From these results, it is, therefore, inconclusive whether the model is tending very slowly to RMT statistics (as favored by Fig.~\ref{fig:LAC_finite_size}-$(a)$) or if it follows some type of intermediate statistics. Considerably larger sector dimensions are, unfortunately, out of reach of current computational capabilities.

\textit{Model C.---}
Finally, we consider a chaotic Liouvillian, model C. Here, the universal limit of RMT statistics is quickly attained. Figure~\ref{fig:LAC_finite_size} depicts the two single-number signatures $\av{\cos\theta}$ and $\av{r}$ and confirms the fast convergence. For the largest sectors diagonalized, the results are already compatible, within their statistical errors, with the infinite-size limit.

\subsection{Disordered open system and detection of many-body localized regime}
\label{subsection:examples_MBL}

After a quench, local observables of chaotic systems thermalize to values that can be predicted by a thermodynamic ensemble average~\cite{vidmarRigol2016}. 
However, in the presence of sufficiently strong disorder for a given system size
\footnote{Recent results~\cite{suntajs2019}
based on the study of spectral fluctuations suggest that the crossover disorder strength beyond which the system is localized is extensive in the system size},
isolated quantum systems, even interacting ones, may fail to thermalize---a phenomenon dubbed many-body localization (MBL)~\cite{nandkishore2015,abanin2019}.
Spectral properties in the many-body localized regime resemble those of integrable models. In fact, some proposals to model MBL rely on approximate locally conserved quantities~\cite{chandran2015}.  
Recently, numerical observation of the MBL regime has also been reported for non-Hermitian Hamiltonians~\cite{hamazaki2018}. 
Moreover, within the delocalized (ergodic) regime, we show that the complex spacing ratio distribution is able to distinguish between GinUE statistics and those of another symmetry class, AI$^\dagger$~\cite{kawabata2018ST,hamazaki2019}. 
This result firmly supports the claim of Ref.~\cite{hamazaki2018} that the model considered therein belongs to this symmetry (universality) class.
 
\subsubsection{Model}
We consider the model of Ref.~\cite{hamazaki2018} consisting of hard-core bosons on a one-dimensional lattice with $N$ sites and periodic boundary conditions. Non-Hermiticity arises due to an alternating on-site gain/loss terms. The (non-Hermitian) Hamiltonian reads
\begin{equation}\label{eq:H_noTRS}
\begin{split}
    H=\sum_{j=1}^N\bigg[
    &-J\left(b_{j+1}^\dagger b_j+b_j^\dagger b_{j+1}\right)\\
    &+Un_jn_{j+1}+\left(h_j+i(-1)^j\gamma\right) n_j\bigg]\,,
\end{split}
\end{equation}
where $b_j^\dagger$ ($b_j$) is the creation (annihilation) operator of a hard-core boson at site $j$, $n_j=b_j^\dagger b_j$ is the particle-number operator, $J$ is the hopping strength, $U$ gives short-range repulsion, $\gamma$ measures the non-Hermiticity, and the local disorder $h_j$ is uniformly distributed in $[-h,h]$. The Hamiltonian conserves particle number, hence we divide the Hilbert space into sectors of fixed particle number $D$. We decompose $H=\bigoplus_{D=0}^N H_D$, where each $H_D$ is a $\binom{N}{D}\times\binom{N}{D}$ matrix. 

\subsubsection{Numerical results}

\begin{figure}[tbp]
    \centering
    \includegraphics[width=0.99\columnwidth]{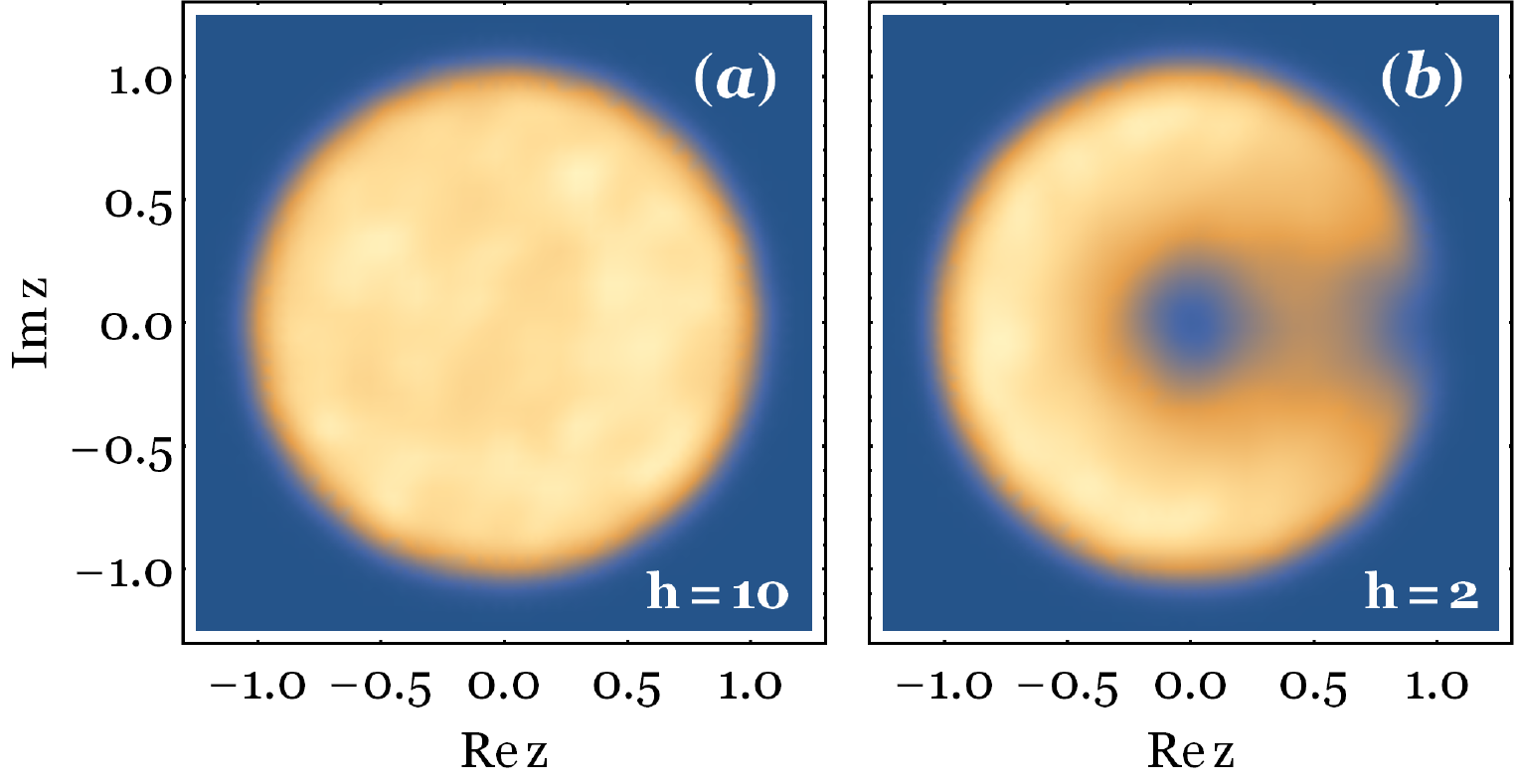}
    \caption{Complex spacing density for the non-Hermitian Hamiltonian of Eq.~(\ref{eq:H_noTRS}), in the $(a)$ localized and $(b)$ delocalized regime, for $N=18$, $D=9$.}
    \label{fig:MBL_ratio}
\end{figure}

\begin{figure}[tbp]
    \centering
    \includegraphics[width=0.99\columnwidth]{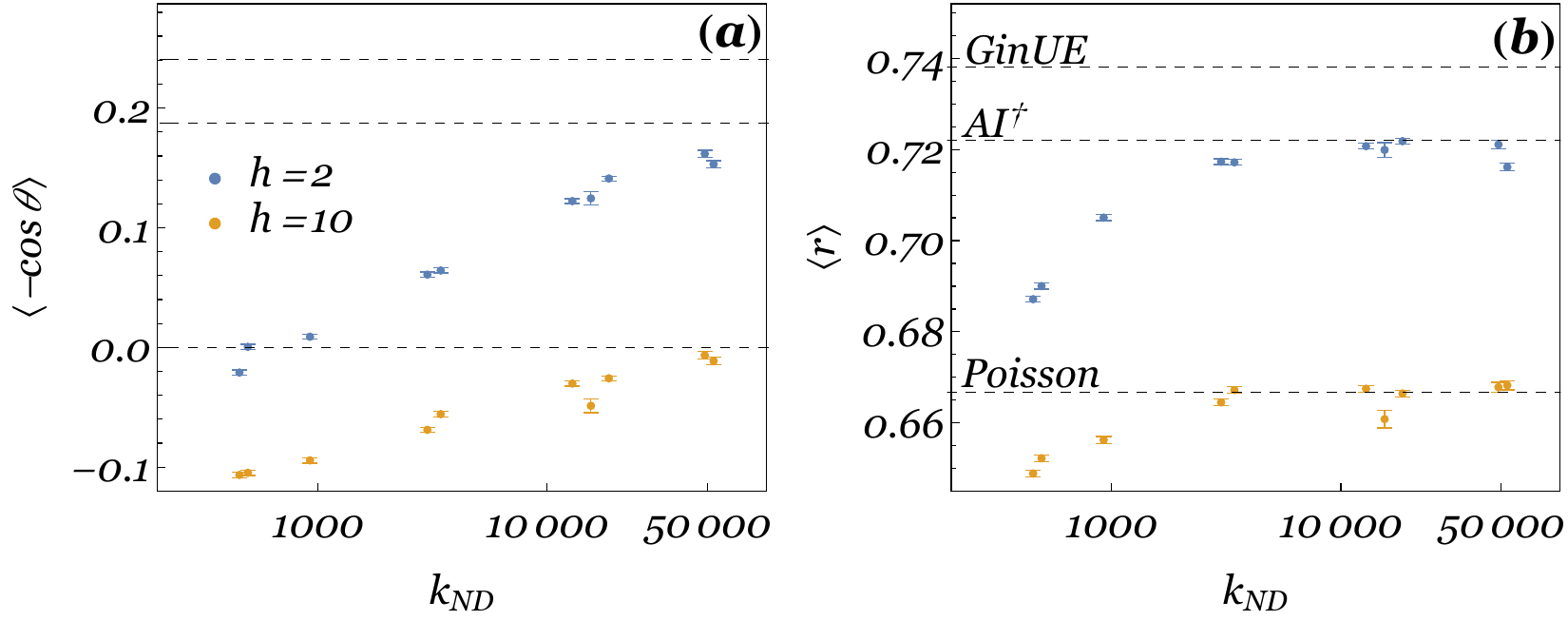}
    \caption{Finite-size effects on the complex spacing ratios of the non-Hermitian Hamiltonian of Eq.~(\ref{eq:H_noTRS}), in the delocalized (blue) and localized (orange) regimes. We consider different chain lengths $N$ and particle-number sectors $D$, the sector dimension being $k_{N\!D}=\binom{N}{D}$. $(a)$: average value of $\cos\theta$; $(b)$: average value of $r$. The upper, middle, and lower dashed lines correspond to the large-$N$ GinUE-, AI$^\dagger$- and Poisson-limits, respectively.}
    \label{fig:MBL_noTRS_finite_size}
\end{figure}

Following Ref.~\cite{hamazaki2018}, we set $J=1$, $U=2$, $\gamma=0.1$ (weak non-Hermiticity) and consider $h=2$ (corresponding to delocalized regime) and $h=10$ (localized regime) separately. Again, numerical results were obtained by exact diagonalization of the Hamiltonian of Eq.~(\ref{eq:H_noTRS}) in sectors of definite particle number $D$. We considered different filling fractions $\nu=D/N$, $\nu=1/2,1/3,1/5$ and system sizes $N$. We performed disorder averaging, obtaining at least $10^5$ eigenvalues for each combination of $N,D$. The largest system diagonalized was for $N=25$, $D=5$, which corresponds to $53130\times53130$ matrices. 

Applying the numerical procedure described at the start of Sec.~\ref{sec:main_results}, we computed the distribution of the complex spacing ratios for the localized and delocalized regimes, see Fig.~\ref{fig:MBL_ratio}. For the system sizes considered there ($N=18,D=9$), the localized regime, Fig.~\ref{fig:MBL_ratio}-$(a)$, supports Poisson statistics (flat distribution), while the delocalized regime, Fig.~\ref{fig:MBL_ratio}-$(b)$, conforms to RMT statistics. These considerations are put on more quantitative grounds by considering single-number signatures for both regimes, see Fig.~\ref{fig:MBL_noTRS_finite_size}, where we plot the values of $\av{\cos\theta}$ and $\av{r}$ for different system sizes and sectors. 

While in the localized regime ($h=10$) the finite-size scaling is consistent with a statistical signature of uncorrelated levels, the delocalized regime, even if clearly non-Poissonian, does not conform to GinUE statistics.  
Instead, it attains the values labeled by AI$^\dagger$, obtained by sampling random matrices from the AI$^\dagger$ symmetry class~\cite{kawabata2018ST,hamazaki2019}. 

These findings show that complex spacing ratios are not only effective in discriminating between localized and delocalized phases, but they can also be used to distinguish between random matrix ensembles with different symmetries.  

\subsection{Classical stochastic process}
\label{subsection:examples_classical_Markov}
Classical stochastic processes are widely used to model physical, chemical and biological systems. 
The solution for a classical stochastic process is obtained by specifying the continuous-time evolution of a probability vector of the system, $\vec{P}$, governed by a Markov matrix $M$: $\partial_t \vec{P}(t)=M\vec{P}(t)$, i.e.\ $\vec{P}(t)=\exp{M t}\vec{P}(t=0)$. By conservation of probability, the columns of the Markov matrix must add up to zero. It then follows that the diagonal elements of $M$ are fully determined by the off-diagonal elements and we can write $M_{jk}=A_{jk}-\delta_{jk}\sum_{m}A_{mk}$, with $\delta_{jk}$ the Kronecker delta and $A_{jj}=0$.
Among the most-studied classical stochastic process are
asymmetric simple exclusion processes (ASEP), used to study transport of interacting particles in one dimension. 
In the following, we analyze the complex-valued spectrum of the matrix $M$ for integrable and nonintegrable ASEP using the complex spacing ratio distribution. We show that while the first case follows Poisson statistics of uncorrelated levels, the second conforms to RMT predictions. 

\subsubsection{Model}
Consider a set of hard-core classical particles on an $N$-site ring with nearest neighbor hoppings. 
The hard-core condition reduces the dimension of configuration space to $2^N$.
Within each time interval $\d t$, any particle can hop from site $j$ to site $j+1$ with probability $p\d t$ and from site $j$ to site $j-1$ with probability $q \d t$. When $p\neq q$, this case defines the  ASEP~\cite{derrida1998,mallick2011,schutz2001,krapivsky2010}.
To break integrability, we further consider a staggering of the hoping probabilities by requiring that the probability of hopping from odd to even sites ($p_1 \d t$ if hopping clockwise) is different from that of hopping from even to odd sites ($p_2 \d t$), and similarly from anticlockwise jumps, with probabilities $q_1 \d t$ and $q_2 \d t$, respectively. Finally, we admit the possibility of particles entering or leaving the system at site $j=1$, with probabilities $\mu_+\d t$ and $\mu_- \d t$, in each time interval.
Assuming $N$ to be even, the matrix $A$ for this process is given by
\begin{equation}\label{eq:Markov_matrix}
\begin{split}
    A=\sum_{j=1}^{N/2}\Big[&\,
    p_1\,\sigma^-_{2j-1}\sigma^+_{2j}+
    p_2\,\sigma^-_{2j}\sigma^+_{2j+1}+
    q_1\,\sigma^+_{2j-1}\sigma^-_{2j}\\
    &+q_2\,\sigma^+_{2j}\sigma^-_{2j+1}
    \Big]+
    \mu_+\sigma^+_1+
    \mu_-\sigma^-_1.
\end{split}
\end{equation}

\begin{figure}[tbp]
    \centering
    \includegraphics[width=0.99\columnwidth]{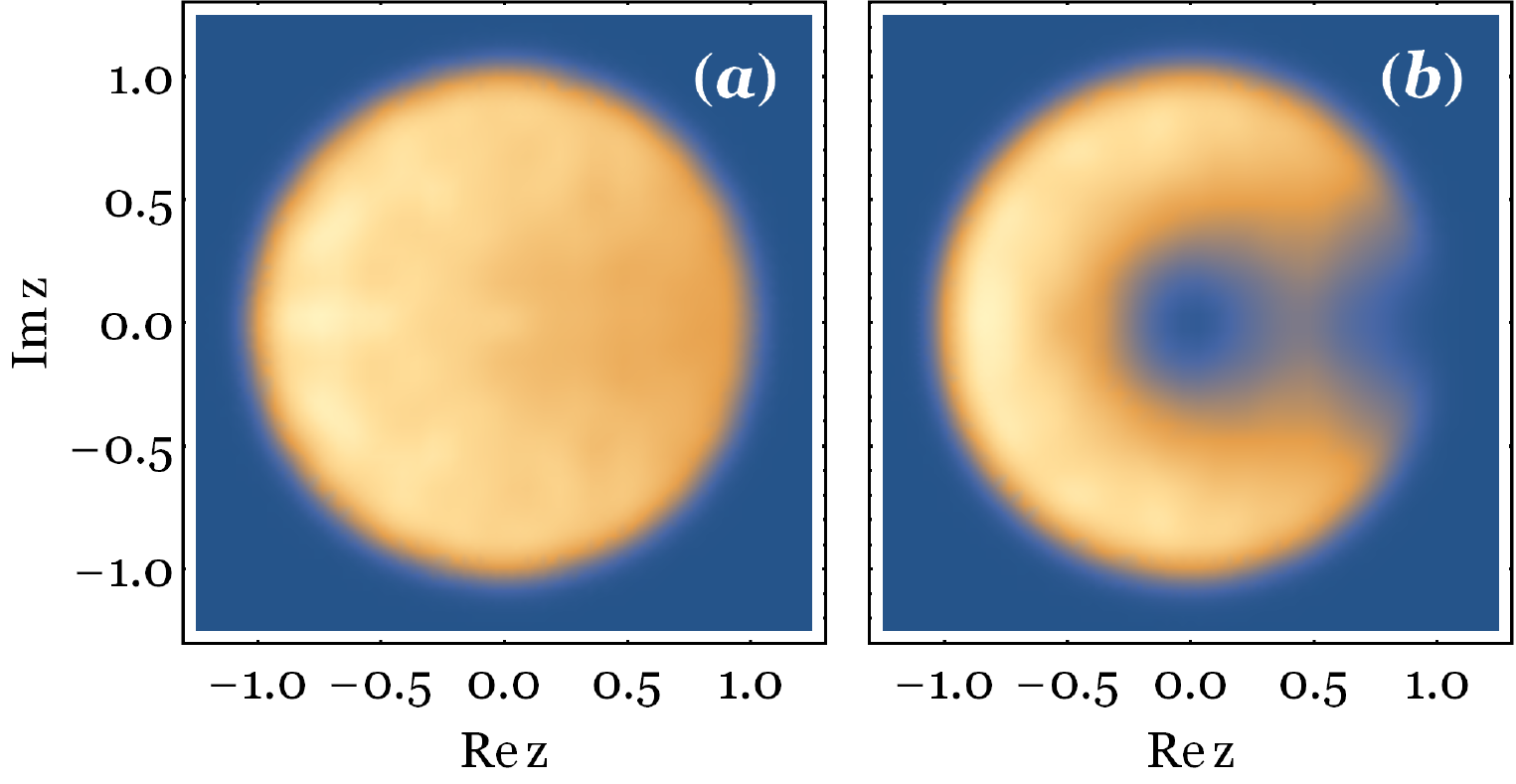}
    \caption{Complex spacing density for the Markov matrix describing the ASEP, with $(a)$ non-staggered and $(b)$ staggered hopping probabilities, for $N=16$.}
    \label{fig:ASEP_ratio}
\end{figure}

\subsubsection{Numerical results}
We numerically diagonalized the Markov matrix $M$, described in the preceding section, for a ring with $N=16$ sites ($M$ is $65536\times65536$). For simplicity, we considered a totally asymmetric exclusion process (TASEP, $q_1=q_2=0$), fixed $p_2=1$ and set $\mu_+=\mu_-=0.5$. Since $\mu_+,\mu_-\neq0$, particle conservation is broken, and hence, we do not restrict $M$ to sectors of fixed particle number. We then considered two cases: non-staggered hopping, $p_1=p_2=1$, for which the model is known to be integrable~\cite{derrida1998}, and staggered hopping, $p_1=0.2\neq p_2$, which we expect to break integrability. This expectation is confirmed by the distribution of complex spacing ratios, see Fig.~\ref{fig:ASEP_ratio}. The complex spacing ratio distribution for non-staggered hopping, Fig.~\ref{fig:ASEP_ratio}-$(a)$ is approximately flat (the inhomogeneity can, as before, be related to finite-size effects). The distribution for staggered hopping, Fig.~\ref{fig:ASEP_ratio}-$(b)$ clearly presents level repulsion and suppression at small angles, with $\av{-\cos\theta}=0.2356(24)$ and $\av{r}=0.7382(7)$. Both effects are compatible with the Ginibre universality class (recall that, for $10^4\times10^4$ matrices from the GinUE, we found $\av{-\cos\theta}=0.24051(61)$ and $\av{r}=0.73810(18)$).
These results show that the complex spacing ratio distribution is also capable of discriminating between integrable and nonintegrable classical stochastic processes. 

\section{Conclusions and Outlook}
\label{sec:conclusions}

We introduced complex spacing ratios to analyze universal spectral features of non-Hermitian systems (integrable and chaotic). We found that angular correlations between levels in dissipative systems provide a clean signature of quantum chaos: uncorrelated random variables, which describe integrable systems, have a flat, and hence isotropic, ratio distribution in the complex plane, while for RMT ensembles from the Ginibre universality class there is a suppression of small angles in the large-$N$ limit. We also reencountered the familiar cubic level repulsion in the latter case. 

Our results show that complex spacing ratios allow one to clearly distinguish (known or conjectured) integrable systems from chaotic ones. Compelling numerical evidence for this claim has been given by a finite-size analysis of boundary-driven spin-chain Liouvillians and classical stochastic processes. 
Complex spacing ratios can also differentiate the many-body-localized regime from the delocalized regime in the non-Hermitian disordered many-body systems. Furthermore, in the delocalized phase, single number signatures, $\av{-\cos\theta}$ and $\av{r}$, can also discriminate between Hamiltonians in different symmetry classes. 

We provided surmises of the large-$N$ complex spacing ratio distribution for GUE and GinUE ensembles.
These surmises were obtained for small matrices, with $N=3,4$, using the CUE and its two-dimensional generalization---the Toric Unitary Ensemble---, which overcome the large finite-size effects observed for small-size GUE and GinUE matrices. 
Even so, the angular marginal distribution was found to have a somewhat slow convergence towards the $N\to\infty$ limit.

Because of their ability to unambiguously discriminate between regular and chaotic dynamics, without the need for unfolding, we expect complex spacing ratios to play an important role in future studies of dissipative quantum chaos and classical stochastic processes. Specifically, complex spacing ratio statistics can be used as a clean and simple empirical detector of integrability, as well as an order parameter characterizing ergodicity-breaking transitions in non-Hermitian systems.

An interesting open question is whether complex spacing ration can be used to discriminate between symmetry classes other than the Ginibre and AI$^{\dagger}$, for instance, those introduced recently in Ref.~\cite{hamazaki2019}.

Finally, the toric unitary ensemble introduced in Sec.~\ref{sec:main_results}, modeling the Coulomb gas on the Clifford torus, also warrants further study. Besides analyzing the properties of random matrix realizations of this novel ensemble, it would be interesting to encounter physical systems for which the TUE arises naturally. 

\begin{acknowledgments}
LS acknowledges support by FCT through PhD Scholarship SFRH/BD/147477/2019. PR acknowledges support by FCT through the Investigador FCT contract IF/00347/2014 and Grant No. UID/CTM/04540/2019. TP acknowledges ERC Advanced grant 694544-OMNES and ARRS research program P1-0402.	
\end{acknowledgments}

\appendix
\begin{widetext}

\section{\texorpdfstring{\uppercase{Uncorrelated random variables}}{Uncorrelated random variables}}
\label{sec:poisson}
Taking isotropy as a starting point, i.e.\ assuming that the distribution of complex spacing ratios only depends on the absolute value of the ratio, $r$, we now show that it is, indeed, flat for uncorrelated random variables (the Poisson spectrum). The independence of the levels simplifies the problem enough so that we are also able to exactly compute the more general ratio, $r_{mk}$, of the distance to the $m^\mathrm{th}$-nearest neighbor ($m$NN) by the distance to the $k^\mathrm{th}$-nearest neighbor ($k$NN) (which reduces to the ratio discussed in the main text when $m=1$, $k=2$). We do all calculations for spectral points (levels) represented by vectors in $d$-dimensional Euclidean space. Real, complex, and quaternionic spectra correspond to $d=1,2,4$, respectively, but our results also apply to other cases, say, uncorrelated random vectors in three-dimensional space.

\subsection{Joint spacing distributions}
\label{subsection:ratio_joint_spacings}
By translational invariance, we can consider the level for which the ratio is being computed (the reference level) at the origin. To compute the probability $\hat{P}(s)\d s$ of finding its NN at a distance $s$, we introduce the conditional probability $g(s)\d s$ of finding a level in $[s,s+\d s]$ given our reference level at the origin, and the probability $H(s)=\int_s^\infty\d s'\hat{P}(s')=1-\int_0^s\d s'\hat{P}(s')$ of having no level in $[0,s]$ (the hole probability). By independence of the levels, the probability $g(s)\d s$ is actually independent of the presence of the reference level. For the NN to be at $s$ we must verify that \textit{(i)} there is a level at $s$ and \textit{(ii)} there are no levels in $[0,s]$, whence we conclude that
\begin{equation}\label{eq:P=gH}
    \hat{P}(s)=g(s)\,H(s)\,.
\end{equation}
Noting that the hole probability is equal to $1-F(s)$, where $F(s)$ is the cumulative distribution of $\hat{P}(s)$, we can equally well express $\hat{P}(s)$ solely in terms of $H(s)$, $\hat{P}(s)=-\frac{\d H}{\d s}$. Alternatively, we can also write $g(s)$ as a function of $H(s)$ only, $g(s)=-\frac{1}{H}\frac{\d H(s)}{\d s}=-\frac{\d\log H}{\d s}$, or, inverting this relation, $H(s)\propto \exp{-\int_0^s\d s'g(s')}$. Finally, this process allows us to express $\hat{P}(s)$ solely in terms of $g(s)$ as $\hat{P}(s)\propto g(s)\,\exp{-\int_0^s\d s'g(s')}$, or, after inverting, $g(s)=\hat{P}(s)/\int_{rs}^\infty\d s'\, \hat{P}(s')$.

Now, Eq.~(\ref{eq:P=gH}) is easily generalized to give the joint distribution of the NN- and NNN-spacing (i.e.\ of the probability density $\hat{P}(s_1,s_2)$ of having the NN at a distance $s_1$ and the NNN at a distance $s_2$), which we need to compute the distribution of their ratio. It is given by considering one level each at $s_1$ and $s_2$ and all remaining levels beyond $s_2$, i.e.
\begin{equation}\label{eq:NN_NNN_joint_spacing}
    \hat{P}(s_1,s_2)=g(s_1)\,g(s_2)\heav{s_2-s_1}H(s_2).
\end{equation}
Analogously, the joint distribution of the first-$k$NN spacings is
\begin{equation}\label{eq:kNN_joint_spacings}
    \hat{P}(s_1,\dots,s_k)=\prod_{j=1}^kg(s_j)\heav{s_{j+1}-s_j}H(s_k)\,.
\end{equation}

It is worthwhile to note that we can express the whole hierarchy of joint probabilities solely in terms of the single-variable functions $\hat{P}(s)$, $g(s)$ or $H(s)$, whichever is easier to compute in a given situation. Of course, this factorization property is a particularity of independent random variables, and does not carry over to random matrix ensembles.

The distribution of the (absolute value) of the ratio $r=s_1/s_2$ is given in terms of the joint distribution $\hat{P}(s_1,s_2)$, and, hence, it is also completely determined by the single spacing distribution $\hat{P}(s)$ [Eq.~(\ref{eq:P(r)_definition})]:
\begin{equation*}
    \varrho(r)=\int \d s_1\d s_2\, \hat{P}(s_1,s_2)\dirac{r-\frac{s_1}{s_2}}
    =\int \d s\, s\, \hat{P}(rs,s)
    =\Theta(1-r)\int_0^\infty \d s\, \frac{s\,\hat{P}(s)\,\hat{P}(rs)}{\int_{rs}^\infty\d s'\, \hat{P}(s')}\,.
\end{equation*}
In the last line, we have expressed the ratio distribution solely in terms of the single spacing probability. Now, we only need to compute one of $\hat{P}(s)$, $g(s)$, or $H(s)$, which we do in $d$ dimensions in the next section.

Likewise, the $m$NN by $k$NN ratio, $r_{mk}\equiv s_m/s_k$, is defined in terms of the joint spacing distribution $\hat{P}\left(s_1,\dots,s_k\right)$ and is fully determined by the single spacing distribution:
\begin{equation}\label{eq:ratio_r_mk_def}
\begin{split}
    \varrho_{mk}&\left(r_{mk}\right)
    =\int\d s_1\cdots\d s_m\cdots \d s_k\,\hat{P}\left(s_1,\dots,s_m,\dots,s_k\right)\dirac{r-\frac{s_m}{s_k}}\\
    &=\int\d s_1\cdots\d s_{m-1}\d s_{m+1}\cdots \d s_k\,\hat{P}\left(s_1,\dots,s_{m-1},rs_k,s_{m+1},\dots,s_k\right)\\
    &=\heav{1-r}\int_0^\infty\d s_1\cdots\d s_{m-1}\d s_{m+1}\cdots \d s_k\heav{s_{k}-s_{k-1}}\cdots\heav{s_{m+1}-rs_k}\heav{rs_k-s_{m-1}}\cdots\heav{s_2-s_1}\\
    &\times\frac{s_k\hat{P}\left(s_1\right)\cdots\hat{P}\left(s_{m-1}\right)\hat{P}\left(rs_k\right)\hat{P}\left(s_{m+1}\right)\cdots\hat{P}\left(s_k\right)}{\int_{s_1}^\infty\d s_1'\hat{P}\left(s'_1\right)\cdots\int_{s_{m-1}}^\infty\d s_{m-1}'\hat{P}\left(s'_{m-1}\right)\int_{rs_k}^\infty\d s'\hat{P}\left(s'\right)\int_{s_{m+1}}^\infty\d s_{m+1}'\hat{P}\left(s'_{m+1}\right)\cdots\int_{s_{k-1}}^\infty\d s_{k-1}'\hat{P}\left(s'_{k-1}\right)}.
\end{split}
\end{equation}

\subsection{Uncorrelated random variables in \texorpdfstring{$d$}{d}-dimensional space}
We consider the spectrum to be composed of $N$ iid random variables, supported in a $d$-dimensional ball of radius $R$. At a later point, we take the limits $N,R\to\infty$ with constant mean density $NR^{-d}=1$. The probabilities $g(s)\d s$ and $H(s)$ are then given by ratios of $d$-dimensional volumes $V_d(L)=\pi^{d/2}/\Gamma(d/2+1)L^d$, where $L$ is a length.

To determine $g(s)$, we note that any one of the $N-1$ levels can be the NN if it falls inside the interval $[s,s+\d s]$, whence it follows that
\begin{equation}\label{eq:Poi_g_finiteN}
\begin{split}
    g(s)\d s&=(N-1)\frac{V_d(s+\d s)-V_d(s)}{V_d(R)}
    =\frac{N-1}{R^d}d\,s^{d-1}\d s+\mathcal{O}(\d s^2)\,.
\end{split}
\end{equation}
Taking the limits $N,R\to\infty$, we immediately obtain that $g(s)\propto s^{d-1}$.

Regarding $H(s)$, since all other $(N-2)$ levels are independent and must lie beyond a distance $s$, we have
\begin{equation}\label{eq:Poi_H_finiteN}
    H(s)=\left(1-\frac{V_d(s)}{V_d(R)}\right)^{N-2}=\left(1-\frac{s^d}{R^d}\right)^{N-2}\,.
\end{equation}

To be able to properly take the limits, we need to unfold the spectrum to a unit mean, i.e.\ we change variables to $\mathfrak{s}=s/\langle s\rangle$. Note that, in the computation of $g(s)$, the unfolding would only give an overall constant, so we did not need it to proceed. Using Eqs.~(\ref{eq:P=gH}), (\ref{eq:Poi_g_finiteN}), and (\ref{eq:Poi_H_finiteN}), we have
\begin{equation}
    \hat{P}(s)=d\,\frac{N-1}{R^d}s^{d-1}\left(1-\frac{s^d}{R^d}\right)^{N-2}\,,
\end{equation}
which is correctly normalized, as it should be. We then have 
\begin{equation}\label{eq:<s>}
    \langle s \rangle=\int_0^\infty\d s\, s\, \hat{P}(s)=\Gamma(1+1/d)\frac{\Gamma(N)}{\Gamma(N+1/d)}R\,,
\end{equation}
or, taking $N\to\infty$ and using the asymptotic behavior of the $\Gamma$ function, $\lim_{N\to\infty}N^\alpha\Gamma(N)/\Gamma(N+\alpha)=1$, for any $\alpha\in\mathbb{C}$, $\langle s \rangle=\Gamma(1+1/d)N^{-1/d}R$.

In terms of the unfolded variable $\s$, the hole probability reads
\begin{equation}\label{eq:Poi_unfolded_H}
  H(\s)=\left(1-\frac{\Gamma(1+1/d)^d\,\s^d}{N}\right)^{N-2}\,.
\end{equation}
Taking limits, $H(\s)=\exp{-\Gamma(1+1/d)^d\,\s^d}$
and the (unfolded) spacing distribution is given by
\begin{equation}\label{eq:Poi_unfolded_P}
    \hat{P}(\s)=d\,\Gamma(1+1/d)^d\s^{d-1}e^{-\Gamma(1+1/d)^d\,\s^d}\,.
\end{equation}
Note that, for $d=1$, we recover the standard exponential distribution, $\hat{P}(\s)=e^{-\s}$. In $d$ dimensions, the spacing follows, instead, a Brody distribution~\cite{brody1973}.

The NN- and NNN-joint spacing distribution $\hat{P}\left(\s_1,\s_2\right)$ can be written solely in terms of the single spacing distribution $\hat{P}(\s)$ by inserting Eq.~(\ref{eq:Poi_unfolded_P}) into Eq.~(\ref{eq:NN_NNN_joint_spacing}),
\begin{equation}\label{eq:joint_NN_NNN_ddim}
\begin{split}
    \hat{P}\left(\s_1,\s_2\right)
    &=d^2\,\Gamma(1+1/d)^{2d}\s_1^{d-1}\s_2^{d-1}
    \times e^{-\Gamma(1+1/d)^d\s_2^d}\heav{\s_2-\s_1}.
\end{split}
\end{equation}

Finally, the joint distribution of the $k$NN spacings, Eq.~(\ref{eq:kNN_joint_spacings}), reads, in $d$ dimensions:
\begin{equation}\label{eq:joint_k_ddim}
\begin{split}
    \hat{P}(\s_1,\dots,\s_k)&=d^k\,\Gamma(1+1/d)^{k d}\prod_{j=1}^{k-1}\s_j^{d-1}e^{-\Gamma(1+1/d)^d\s_k^d}
    \times\prod_{j=1}^k\Theta(\s_{j+1}-\s_j)\,.
\end{split}
\end{equation}

\subsection{Ratio distribution}

We now turn to the ratio distributions. Henceforth, we always assume that we are at the unfolded scale and denote the spacings by $s$ instead of $\s$.
Inserting Eq.~(\ref{eq:Poi_unfolded_P}) into the last equality of Eq.~(\ref{eq:P(r)_definition}), we obtain Eq.~(\ref{eq:NN_NNN_ratio_r_ddim}),
\begin{equation*}
        \varrho(r)=d\, r^{d-1}\heav{1-r}.
\end{equation*}
The constraint enforced by the $\Theta$-function implies that the distribution is supported in the $d$-dimensional unit ball, which we parametrize by the radial distance $r$ and the $(d-1)$-dimensional solid angle $\Omega_{d-1}$. By recalling that $\int_0^1\d r\, r^{d-1}\int\d\Omega_{d-1}\varrho\left(r,\Omega_{d-1}\right)=\int_0^1\d r\varrho(r)$ and that the distribution is isotropic and hence $\varrho(r,\Omega_{d-1})$ is independent of $\Omega_{d-1}$, by using Eq.~(\ref{eq:NN_NNN_ratio_r_ddim}), and by noting that $\int\d\Omega_{d-1}=S_{d-1}$ gives the area of the unit sphere in $d$ dimensions and that $S_{d-1}/V_d(1)=d$, we conclude that
\begin{equation}\label{eq:poisson_ddim_ratio_final}
    \varrho(r,\Omega_{d-1})=\heav{1-r}\frac{d}{S_{d-1}}=\heav{1-r}\frac{1}{V_d(1)},
\end{equation}
i.e.\ the distribution is indeed flat since it is given by the inverse of the volume of its support. 

We next consider the distribution of the $m$NN by $k$NN ratio, of which the above result is a special case ($m=1$, $k=2$). Inserting Eq.~(\ref{eq:Poi_unfolded_P}) into the last equality of Eq.~(\ref{eq:ratio_r_mk_def}), we obtain
\begin{equation}\label{eq:ratio_r_mk_final}
    \varrho_{mk}(r_{mk})=\binom{k-1}{m}d\,m\,(r_{mk})^{dm-1}(1-(r_{mk})^d)^{k-m-1}\,.
\end{equation}

\begin{figure*}[tbp]
    \centering
    \includegraphics[width=\textwidth]{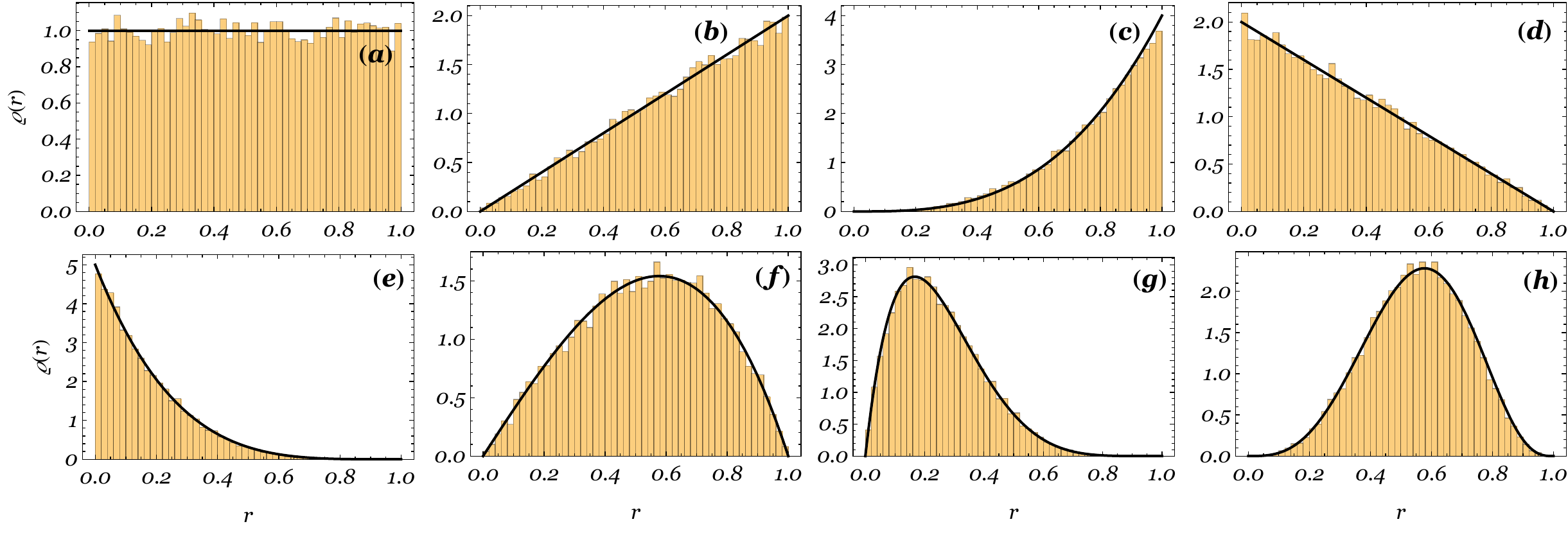}
    \caption{Comparison of analytic prediction for the $m$NN by $k$NN spacing ratio, Eq.~(\ref{eq:ratio_r_mk_final}), black line, with numerical results. Each histogram is obtained by computing the ratios for $20\,000$ iid levels. The numerical parameters are as follows: $(a)$ $d=1$, $m=1$, $k=2$; $(b)$ $d=2$, $m=1$, $k=2$; $(c)$ $d=4$, $m=1$, $k=2$; $(d)$ $d=1$, $m=1$, $k=3$; $(e)$ $d=1$, $m=1$, $k=6$; $(f)$ $d=2$, $m=1$, $k=3$; $(g)$ $d=1$, $m=2$, $k=8$; $(h)$ $d=2$, $m=2$, $k=6$. $(a)$, $(b)$, $(c)$ correspond to the NN-by-NNN spacing ratio, Eq.~(\ref{eq:NN_NNN_ratio_r_ddim}), for real, complex and quaternionic spectra, respectively.}
    \label{fig:analytics_Poisson}
\end{figure*}

Equation~(\ref{eq:ratio_r_mk_final}) constitutes the most general distribution for Poisson spacing ratios in $d$ dimensions. Figure~\ref{fig:analytics_Poisson} shows a comparison of Eq.~(\ref{eq:ratio_r_mk_final}) with numerical spacing ratios of $20\,000$ iid levels, for eight different combinations of $d,m,k$, showing perfect agreement in all cases.

Finally, note that, a flat distribution in the $d$-dimensional unit ball is only possible if $\varrho(r_{mk})\propto (r_{mk})^{d-1}$ for all $r_{mk}$, as in Eq.~(\ref{eq:NN_NNN_ratio_r_ddim}). This case implies $m=1$ and $k-m=1\Rightarrow k=2$. We thus see that a flat distribution is a peculiarity of the NN-by-NNN ratio and is not achieved by any other combination of $m,k$. 


\section{\texorpdfstring{\uppercase{Hermitian random matrix ensembles}}{Hermitian random matrix ensembles}}
\label{sec:analytics_real}

\subsection{Arbitrary Hermitian ensembles}
We now consider the NN-by-NNN spacing ratio for random matrices, first addressing the case of Hermitian ensembles. Let again the spectrum of an (arbitrary for now) Hermitian-RMT matrix be composed of levels $\{\lambda_k\}_{k=1}^N$, which are \emph{not} taken to be ordered. Since the joint eigenvalue distribution function $P^{(N)}(\{\lambda_k\})$ is invariant under permutations of levels, we reorder the set such that our reference level is $\lambda_1$, its NN $\lambda_2$ and its NNN $\lambda_3$. Contrary to the previous section, in general, we cannot set the reference level $\lambda_1=0$ since $P^{(N)}(\{\lambda_k\})$ may not be invariant under translations. These choices (together with the immediate implication that all other $N-2$ levels must be further away from $\lambda_1$ than $\lambda_3$ is) are enforced via the constraint
\begin{equation}
\begin{split}
    &\heav{(\lambda_3-\lambda_1)^2-(\lambda_2-\lambda_1)^2}\prod_{j>3}^{N}\heav{(\lambda_j-\lambda_1)^2-(\lambda_3-\lambda_1)^2}.
\end{split}
\end{equation}
The NN-by-NNN ratio is $r=(\lambda_2-\lambda_1)/(\lambda_3-\lambda_1)$. We can then immediately write down the expression for its distribution,
\begin{equation}
    \varrho^{(N)}(r)=\int\d\lambda_1\cdots\d\lambda_NP^{(N)}(\lambda_1,\dots,\lambda_N)\dirac{r-\frac{\lambda_2-\lambda_1}{\lambda_3-\lambda_1}}
    \heav{(\lambda_3-\lambda_1)^2-(\lambda_2-\lambda_1)^2}\prod_{j>3}^{N}\heav{(\lambda_j-\lambda_1)^2-(\lambda_3-\lambda_1)^2}.
\end{equation}

We next change variables to $u\equiv \lambda_1$, $v\equiv\lambda_3-\lambda_1$, $s_n=\lambda_{n+3}-\lambda_1$ ($n=1,\dots,N-3$), perform the integration in $\lambda_2$ using the $\delta$-function, and obtain
\begin{equation}\label{eq:hermitian_ratio_general}
    \varrho^{(N)}(r)=\heav{1-r^2}\int \d u\d v\prod_{j=1}^{N-3}\d s_j\heav{s_j^2-v^2}\, \abs{v}\,P^{(N)}\left(u,u+rv,u+v,u+s_1,\dots,u+s_{N-3}\right).
\end{equation}

\subsection{Gaussian ensembles}
\label{subsection:ratio_hermitian_gaussian}
Equation (\ref{eq:hermitian_ratio_general}) is valid for an arbitrary Hermitian ensemble. We now specialize for the case of the Gaussian ensembles, GO/U/SE, labeled by the Dyson index $\beta$. The joint eigenvalue distribution reads
\begin{equation}
    P^{(N)}_\mathrm{GE}(x_1,\dots,x_N)\propto\exp{-\frac{1}{2}\sum_{j=1}^Nx_j^2}\prod_{j>k}^{N}\abs{x_j-x_k}^\beta\,.
\end{equation}
In terms of the variables of Eq.~(\ref{eq:hermitian_ratio_general}), we have
\begin{equation}
\begin{split}
    P^{(N)}_\mathrm{GE}&(u,u+rv,u+v,u+s_1,\dots,u+s_{N-3})\propto
    \abs{r}^\beta\abs{1-r}^\beta\abs{v}^{3\beta}\prod_{j=1}^{N-3}\abs{s_k}^\beta\abs{s_k-v}^\beta\abs{s_k-rv}^\beta\prod_{j<k}^{N-3}\abs{s_j-s_k}^\beta\\
    &\times\,\exp{-\frac{1}{2}\left[Nu^2+2u\left((1+r)v+\sum_{j=1}^{N-3}s_j\right)+(1+r^2)v^2+\sum_{j=1}^{N-3}s_j^2\right]}\,.
\end{split}
\end{equation}

The integration in $u$ is Gaussian and can be readily performed, yielding
\begin{equation}
    \int\d u \exp{-\frac{1}{2}\left[Nu^2+2u\left((1+r)r+\sum_{j=1}^{N-3}s_j\right)\right]}\propto\exp{\frac{1}{2N}\left[(1+r)v+\sum_{j=1}^{N-3}s_j\right]^2}\,.
\end{equation}
We finally obtain the distribution of the ratio as an $(N-2)$-fold integral:
\begin{equation}\label{eq:hermitian_ratio_gaussian}
\begin{split}
    \varrho^{(N)}_\mathrm{GE}(r)&\propto\heav{1-r^2}\abs{r}^\beta\abs{1-r}^\beta\int \d v\,\abs{v}^{3\beta+1}\exp{-\frac{1}{2}v^2\left(1+r^2-\frac{(1+r)^2}{N}\right)}\int\prod_{j=1}^{N-3}\d s_j \heav{s_j^2-v^2}\\
    &\times\abs{s_j}^\beta \abs{s_j-v}^\beta \abs{s_j-rv}^\beta\exp{-\frac{1}{2}\left(s_j^2- \frac{1+r}{N}vs_j\right)}
    \exp{-\frac{1}{N}\sum_{k,\ell=1}^{N-3}s_ks_\ell}\prod_{k<\ell}^{N-3}\abs{s_k-s_\ell}^\beta\,.
\end{split}
\end{equation}

For small $N$ ($N=3,4$) the integrals in Eq.~(\ref{eq:hermitian_ratio_gaussian}) can be computed exactly. Unfortunately, contrary to the consecutive spacings ratio, for NN-by-NNN ratios, the small-size expressions do not accurately describe the large-$N$ asymptotics.

For $N=3$, no $s_j$-integrals exist in Eq.~(\ref{eq:hermitian_ratio_gaussian}). Furthermore, the $r$-dependence can be factored out of the $v$-integral and no integrals have to be performed at all:
\begin{equation}\label{eq:gaussian_ratio_N3}
\begin{split}
    \varrho^{(3)}_\mathrm{GE}(r)&\propto\heav{1-r^2}\abs{r}^\beta\abs{1-r}^\beta\int \d v\,v^{3\beta+1}\exp{-\frac{1}{3}v^2\left(1-r+r^2\right)}=\mathcal{N}\frac{\abs{r}^\beta\abs{1-r}^\beta}{(1-r+r^2)^{1+3\beta/2}}\heav{1-r^2},
\end{split}
\end{equation}
where the $\beta$-dependant normalization is $\mathcal N=9/4$ for $\beta=1$, $\mathcal N=27\sqrt{3}/(2\pi)$ for $\beta=2$ and $\mathcal N=243\sqrt{3}/(2\pi)$ for $\beta=4$. The distribution of Eq.~(\ref{eq:gaussian_ratio_N3}) for $\beta=2$ is plotted in Fig.~\ref{fig:analytics_GUE}-$(a)$, in comparison with exact diagonalization results.

For $N=4$, we must perform an additional integral in $s$ (here for $\beta=2$),
\begin{equation}\label{eq:GUE_N=4_start}
\begin{split}
    \varrho^{(4)}_\mathrm{GUE}(r)&\propto\heav{1-r^2}r^2(1-r)^2\int_{-\infty}^{+\infty}\d v\, \abs{v}^7\exp{-\frac{3}{8}v^2\left(1+r^2-\frac{2}{3}r\right)}\\
    &\times\int_{-\infty}^{+\infty}\d s\, s^2(s-v)^2(s-rv)^2\exp{-\frac{3}{8}s^2}\exp{\frac{1}{4}(1+r)vs}\heav{s^2-v^2},
\end{split}
\end{equation}
If we denote 
\begin{equation}
    f(s,v,r)=\int\d s\, s^2(s-v)^2(s-rv)^2\exp{-\frac{3}{8}s^2}\exp{\frac{1}{4}(1+r)vs}\,,
\end{equation}
then Eq.~(\ref{eq:GUE_N=4_start}) reads
\begin{equation}
\begin{split}
     \varrho^{(4)}_\mathrm{GUE}&(r)\propto\heav{1-r^2}r^2(1-r^2)\\
      \times\bigg[
     &\int_0^{+\infty}\d v\abs{v}^7\exp{-\frac{3}{8}v^2\left(1+r^2-\frac{2}{3}r\right)}\left(f(s,v,r)\Big|^{s=-v}_{s=-\infty}+f(s,v,r)\Big|^{s=\infty}_{s=v}\right)\\
    +&\int_{-\infty}^0\d v\abs{v}^7\exp{-\frac{3}{8}v^2\left(1+r^2-\frac{2}{3}r\right)}\left(f(s,v,r)\Big|^{s=v}_{s=-\infty}+f(s,v,r)\Big|^{s=\infty}_{s=-v}\right)\bigg]\,,
\end{split}
\end{equation}
which is evaluated (with the correct normalization) to 
\begin{equation}\label{eq:GUE_N4}
\begin{split}
    \varrho^{(4)}_\mathrm{GUE}(r)=&\,\frac{1}{4\pi}
    \frac{r^2(1-r)^2}{(1-r+r^2)^7(8+3r^2)^{13/2}(4-4r+3r^3)^{9/2}}\\
    &\times\left(\sqrt{8+3r^2}B_1^{(4)}(r)+\sqrt{4-4r+3r^2}B_2^{(4)}(r)+\sqrt{3}\sqrt{8+3r^2}\sqrt{4-4r+3r^2}B_3^{(4)}(r)\right)\heav{1-r^2},
\end{split}
\end{equation}
with the polynomials $B^{(4)}_k$ given in the Supplemental Material~\cite{SM}. 

We compare the analytical predictions for the ratio distribution for small-size matrices with numerical results from exact diagonalization of GUE-drawn random matrices in Fig.~\ref{fig:GUE_CUE_N3_4}-$(a)$ and $(b)$. The agreement is perfect, which was to be expected since the computation is exact. However, these results are not particularly useful in practice  since universality is only displayed for large $N$. We thus turn to the case $N\to\infty$.

\begin{figure*}[tbp]
    \centering
    \includegraphics[width=0.99\columnwidth]{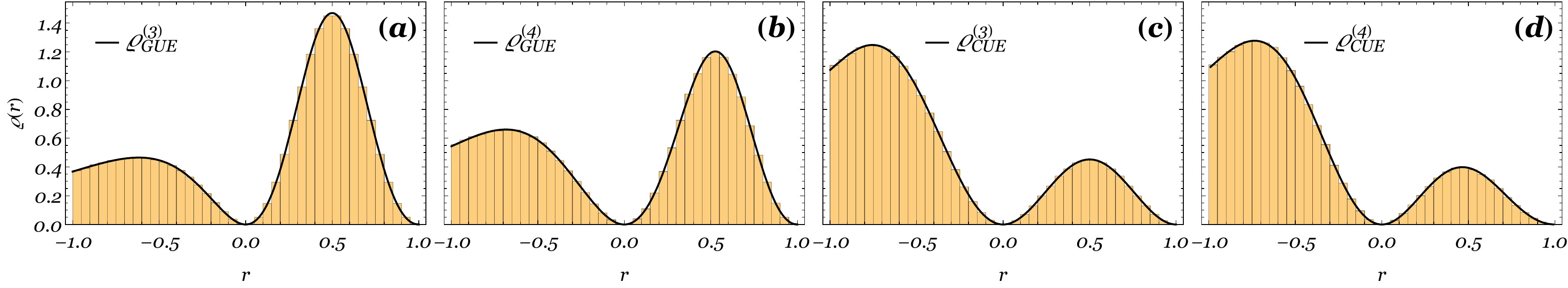}
    \caption{Comparison of exact diagonalization (ED) for GUE- and CUE-drawn random matrices with analytical results. $(a)$: $N=3$ GUE, black line given by Eq.~(\ref{eq:gaussian_ratio_N3}), with $\beta=2$; $(b)$: $N=4$ GUE, black line given by Eq.~(\ref{eq:GUE_N4}); $(c)$: $N=3$ CUE, black line given by Eq.~(\ref{eq:CUE_N3}); $(d)$: $N=4$ CUE, black line given by Eq.~(\ref{eq:CUE_N4}).}
    \label{fig:GUE_CUE_N3_4}
\end{figure*}

When $N\to\infty$, we can rewrite Eq.~(\ref{eq:hermitian_ratio_gaussian}), discarding all exponentials suppressed by $1/N$:
\begin{equation}\label{eq:gaussian_ratio_N_infty}
\begin{split}
    \varrho^{(N\to\infty)}_\mathrm{GE}(r)&\propto\heav{1-r^2}\frac{\abs{r}^\beta\abs{1-r}^\beta}{(1+r^2)^{1+3\beta/2}}\int\d v\, v^{3\beta+1}e^{-v^2}\int\prod_{j=1}^N\d s_j\heav{s_j^2-\frac{v^2}{1+r^2}}\\
    &\times\abs{s_j}^\beta\abs{s_j-\frac{rv}{\sqrt{1+r^2}}}^\beta\abs{s_j-\frac{v}{\sqrt{1+r^2}}}^\beta e^{-s_j^2}\prod_{j<k}^N\abs{s_j-s_k}^\beta.
\end{split}
\end{equation}

Although we cannot compute this integral exactly, note that its multiplying prefactor gives the exact distribution of $r$ for $r\to0$. It also qualitatively describes the distribution for all $r$, albeit it is missing the exact heights of the peaks of positive and negative $r$. We thus obtain the loosely approximating distribution,
\begin{equation}\label{eq:gaussian_ratio_N_infty_approx}
    \varrho^{(N\to\infty)}_\mathrm{GE}(r)\approx\mathcal{N}\frac{r^\beta\abs{1-r}^\beta}{(1+r^2)^{1+3\beta/2}}\heav{1-r^2}.
\end{equation}
At any rate, the absence of a term $-r$ inside the denominator (which was killed by the limit $N\to\infty$) completely distinguishes this result from the case $N=3$, see Fig.~\ref{fig:analytics_GUE} (the black line is the approximation of Eq.~(\ref{eq:gaussian_ratio_N_infty_approx})).

\subsection{Circular ensembles}
\label{subsection:ratio_hermitian_circular}
We discussed in Sec.~\ref{sec:main_results} how the difference between the $N=3$ and $N\to\infty$ statistics is due to boundary effects. To eliminate these effects we should consider periodic boundary conditions, i.e.\ identify the ends of the spectrum. Hence, we consider the circular ensembles, whose spectrum is supported on the unit circle and whose joint eigenvalue distribution is 
\begin{equation}
    P^{(N)}_\mathrm{CE}\left(\phi_1,\dots,\phi_N\right)\propto\prod_{j<k}\abs{e^{i\phi_j}-e^{i\phi_k}}^\beta\,.
\end{equation}

Note that, although the eigenvalues are complex ($e^{i\phi_j}$), they are fully described by real angles $\phi_j\in(-\pi,\pi]$. The spacing ratio is defined in terms of the real variables, $r=(\phi_2-\phi_1)/(\phi_3-\phi_1)$, i.e.\ we are measuring the spacings on the circle, not in the embedding space, $\mathbb{C}$. By rotational invariance of the circle, we may set $\phi_1=0$. 
We can rewrite the Vandermonde interaction as $\abs{e^{i\phi_j}-e^{i\phi_k}}=\sin^\beta(\abs{\phi_j-\phi_k}/2)$. The general result of Eq.~(\ref{eq:hermitian_ratio_general}), applied to the circular ensembles, reads
\begin{equation}\label{eq:circular_ratio_general}
\begin{split}
    \varrho^{(N)}_\mathrm{CE}(r)&\propto\heav{1-r^2}\int_{-\pi}^{\pi} \d v\abs{v}\sin^\beta\frac{\abs{v}}{2}\sin^\beta\frac{\abs{rv}}{2}\sin^\beta\frac{\abs{(1-r)v}}{2}\\
    &\times\int_{-\pi}^{\pi}\prod_j\d s_j\heav{s_j^2-v^2}\sin^\beta\frac{\abs{s_j}}{2}\sin^\beta\frac{\abs{s_j-rv}}{2}\sin^\beta\frac{\abs{s_j-v}}{2}\prod_{j<k}^{N-3}\sin^\beta\frac{\abs{s_j-s_k}}{2}\,.
\end{split}
\end{equation}

We now evaluate the preceding integral for $N=3$ and $N=4$, restricting ourselves to the complex case, $\beta=2$. We have $\sin^2((\phi_j-\phi_k)/2)=2(1-\cos(\phi_k-\phi_j))$.

For $N=3$, a single integral in $v$ is to be performed [Eq.~(\ref{eq:CUE_N3_integral})]
\begin{equation*}
    \varrho^{(3)}_\mathrm{CUE}(r)\propto\heav{1-r^2}\int_{-\pi}^\pi\d v \abs{v}\left(1-\cos v\right)\left(1-\cos rv\right)\left(1-\cos(r-1)v\right),
\end{equation*}
which yields, after normalization,
\begin{equation}\label{eq:CUE_N3}
\begin{split}
    \varrho^{(3)}_\mathrm{CUE}(r)=&\,\frac{1}{48\pi^2}\frac{\heav{1-r^2}}{(r-2)^2(r-1)^2(r-\frac{1}{2})^2r^2(r+1)^2}\\
    \times&\left(Q_1^{(3)}(r)+Q_2^{(3)}(r)\cos(\pi r)+Q_3^{(3)}(r)\cos(2\pi r)+Q_4^{(3)}(r)\sin(\pi r)+Q_5^{(3)}(r)\sin(2\pi r)\right),
\end{split}
\end{equation}
with the polynomials $Q^{(3)}_k(r)$ given in the Supplemental Material~\cite{SM}.
The distribution of Eq.~(\ref{eq:CUE_N3}) is plotted in black in Fig.~\ref{fig:GUE_CUE_N3_4}-$(c)$, in comparison with numerical diagonalization of $N=3$ CUE matrices, and in red in Fig.~\ref{fig:analytics_GUE}-$(b)$, in comparison with diagonalization of large-$N$ GUE matrices.

For $N=4$, we have an additional integral in $s$ to perform,
\begin{equation}\label{eq:CUE_N=4_start}
\begin{split}
    \varrho^{(4)}_\mathrm{CUE}(r)&\propto\heav{1-r^2}\int_{-\pi}^\pi\d v \abs{v}\left(1-\cos v\right)\left(1-\cos rv\right)\left(1-\cos(r-1)v\right)\\
    &\times
    \int_{-\pi}^\pi\d s\left(1-\cos s\right)\left(1-\cos(s-rv)\right)\left(1-\cos(s-v)\right)\heav{s^2-v^2}.
\end{split}
\end{equation}
Following the procedure leading to Eq.~(\ref{eq:GUE_N4}), the integral is evaluated (with the correct normalization) as
\begin{equation}\label{eq:CUE_N4}
\begin{split}
    \varrho^{(4)}_\mathrm{CUE}(r)=&\,\frac{1}{2^{19} 3^{17} \pi ^3}
    \frac{1}{(r-6)^2 (r-5)^2 (r-4)^2 (r-3)^3 \left(r-\frac{5}{2}\right)^2 (r-2)^3 \left(r-\frac{3}{2}\right)^3 \left(r-\frac{4}{3}\right)^2}\\
    \times&\frac{1}{(r-1)^3\left(r-\frac{2}{3}\right)^3 \left(r-\frac{1}{2}\right)^3 \left(r-\frac{1}{3}\right)^3 r^3 \left(r+\frac{1}{3}\right)^2 \left(r+\frac{1}{2}\right)^3 \left(r+\frac{2}{3}\right)^2 (r+1)^3}\\
    \times&\frac{1}{\left(r+\frac{4}{3}\right)^2 \left(r+\frac{3}{2}\right)^2 (r+2)^3\left(r+\frac{5}{2}\right)^2 (r+3)^2 (r+4)^2 (r+5)^2 (r+6)^2}\\
    \times&\Big(Q_1^{(4)}(r)+Q_2^{(4)}(r)\cos(\pi r)+Q_3^{(4)}(r)\cos(2\pi r)+Q_4^{(4)}(r)\cos(3\pi r)\\
    &+Q_5^{(4)}(r)\sin(\pi r)+Q_6^{(4)}(r)\sin(2\pi r)+Q_7^{(4)}(r)\sin(3\pi r)\Big)\,,
\end{split}
\end{equation}
with the polynomials $Q^{(4)}_k$ given in the Supplemental Material~\cite{SM}. The distribution of Eq.~(\ref{eq:CUE_N4}) is plotted in black in Fig.~\ref{fig:GUE_CUE_N3_4}-$(d)$, in comparison with numerical diagonalization of $N=4$ CUE matrices, and in blue in Fig.~\ref{fig:analytics_GUE}-$(b)$, in comparison with exact diagonalization of large-$N$ GUE matrices. Although in the large-$N$ limit we can give only approximate expressions for the complex spacing ratio distribution, the small-size surmises computed for the circular ensembles describe very well the universal large-$N$ asymptotics. Indeed, the distribution for the $N=4$ CUE is already indistinguishable (to the naked eye) from the numerical $N\to\infty$ results.

\section{\texorpdfstring{\uppercase{Non-Hermitian random matrix ensembles}}{Non-Hermitian random matrix ensembles}}
\label{sec:analytics_complex}

\subsection{Arbitrary non-Hermitian ensembles}
We now turn to complex spectra. We consider an arbitrary $N\times N$ matrix from a non-Hermitian ensemble whose complex eigenvalues are $\{\lambda_k\}_{k=1}^{N}$, $\lambda_k=x_k+iy_k$, and their joint distribution function is $P^{(N)}(\lambda_1,\dots,\lambda_N)= P^{(N)}(x_1,\dots,x_N;y_1,\dots,y_N)$.
We again consider the first level $\lambda_1$ to be the reference level, its NN to be $\lambda_2$ and its NNN to be $\lambda_3$. The complex NN-by-NNN spacing ratio is 
\begin{equation}
\begin{split}
    z\equiv&\, r e^{i\theta}\equiv x+iy=\frac{\lambda_2-\lambda_1}{\lambda_3-\lambda_1}=\frac{(x_2-x_1)(x_3-x_1)+(y_2-y_1)(y_3-y_1)}{(x_3-x_1)^2+(y_3-y_1)^2}+i\frac{(x_3-x_1)(y_2-y_1)-(x_2-x_1)(y_3-y_1)}{(x_3-x_1)^2+(y_3-y_1)^2}\,.
\end{split}
\end{equation}
We introduce new variables $u\equiv x_1$, $v\equiv y_1$, $p\equiv x_2-x_1$, $q\equiv y_2-y_1$, $s\equiv x_3-x_1$, $t\equiv y_3-y_1$, $a_n\equiv x_{n+3}-x_1$, $b_n\equiv y_{n+3}-y_1$, $n=1,\dots,{N-3}$. In terms of these new variables, the $\delta$-function constraints (fixing the real and imaginary parts of $z$) are
\begin{equation}\label{eq:delta_constraints}
\begin{split}
    \dirac{x-\frac{ps+qt}{s^2+t^2}}\dirac{y-\frac{sq-pt}{s^2+t^2}}=(s^2+t^2)\,\delta\Big(p-\left(sx-ty\right)\Big)\,\delta\Big(q-\left(tx+sy\right)\Big)\,,
\end{split}
\end{equation}
and the $\Theta$-function constraints (requiring all $\lambda_n$ with $n>3$ to be further away from $\lambda_1$ than $\lambda_3$) are
\begin{equation}\label{eq:theta_constraints}
    \heav{(s^2+t^2)-(p^2+q^2)}\prod_{j=4}^{N}\\\heav{(a_j^2+b_j^2)-(s^2+t^2)}.
\end{equation}

The distribution function of $z$ is again obtained by integrating the joint eigenvalue distribution multiplied by the constraints of Eqs.~(\ref{eq:delta_constraints}) and (\ref{eq:theta_constraints}). Integrating over $p$ and $q$ using the $\delta$-functions, we arrive at the distribution for $z$ for an arbitrary non-Hermitian ensemble:
\begin{equation}\label{eq:nonhermitian_ratio_general}
\begin{split}
    \varrho^{(N)}(x,y)&=\heav{1-(x^2+y^2)}\int\d u\d v\d s\d t\prod_{j=1}^{N-3}\d a_j \d b_j\heav{(a_j^2+b_j^2)-(s^2+t^2)}\,(s^2+t^2)\\
    &\times P^{(N)}(u,u+sx-ty,u+s,u+a_1,\dots,u+a_{N-3};v,v+tx+sy,v+t,v+b_1,\dots,v+b_{N-3})\,.
\end{split}
\end{equation}

\subsection{Ginibre Unitary Ensemble}
\label{subsection:ratio_nonhermitian_ginibre}
We now restrict ourselves to the GinUE (complex Gaussian iid entries, $\beta=2$), whose joint eigenvalue distribution reads:
\begin{equation}
    P^{(N)}_\mathrm{GinUE}(x_1,\dots,x_N;y_1,\dots,y_N)=\prod_{j<k}\left[(x_j-x_k)^2+(y_j-y_k)^2\right]\exp{-\sum_{j=1}^N\left(x_j^2+y_j^2\right)}\,.
\end{equation}

Replacing the $x_j$, $y_j$ by the variables of Eq.~(\ref{eq:nonhermitian_ratio_general}) and performing the Gaussian integration over the two variables $u$, $v$, we arrive at the ratio distribution for the Ginibre ensemble,
\begin{equation}\label{eq:complex_ratio_N}
\begin{split}
    \varrho^{(N)}_\mathrm{GinUE}(x,y)&\propto\heav{1-(x^2+y^2)}(x^2+y^2)(1+x^2+y^2-2x)\\
    &\times\int \d s\d t\,(s^2+t^2)^4\exp{-(s^2+t^2)\left[(1+x^2+y^2)\left(1-\frac{1}{N}\right)-\frac{2}{N}x\right]}\\
    &\times\int\prod_{j=1}^{N-3}\d a_j\d b_j\heav{a_j^2+b_j^2-(s^2+t^2)}\left(a_j^2+b_j^2\right)\left((a_j-s)^2+(b_j-t)^2\right)\\
    &\times\left[(a_j-sx+ty)^2+(b_j-tx-sy)^2\right]\prod_{j<k}^{N-3}\left[(a_j-a_k)^2+(b_j-b_k)^2\right]\\
    &\times\exp{-\sum_{j=1}^{N-3}\left[a_j^2\left(1-\frac{1}{N}\right)-\frac{1}{N}\sum_{k\neq j}a_ja_k-\frac{2}{N}a_j\left\{s(1+x)-t y\right\}\right]}\\
    &\times\exp{-\sum_{j=1}^{N-3}\left[b_j^2\left(1-\frac{1}{N}\right)-\frac{1}{N}\sum_{k\neq j}b_jb_k-\frac{2}{N}b_j\left\{sy+t(1+x\right\}\right]}\,.
\end{split}
\end{equation}

As before, the distribution for $N=3$ follows from Eq.~(\ref{eq:complex_ratio_N}) without the need to perform any integrals explicitly. Indeed, in polar coordinates $x=r\cos\theta$, $y=r\sin\theta$, we get, after normalization,
\begin{equation}\label{eq:GinUE_N3}
    \varrho^{(3)}_\mathrm{GinUE}(r,\theta)=\frac{81}{8\pi}\frac{r^2(1+r^2-2r\cos\theta)}{(1+r^2-r\cos\theta)^5}\heav{1-r}.
\end{equation}
Equation~(\ref{eq:GinUE_N3}) perfectly describes the exact diagonalization results for $N=3$, see Fig.~\ref{fig:GinUE_TUE_N3}.

As for the Hermitian case, the leading order behavior (i.e.\ the first term in a power expansion in $r$) of the distribution for $N\to\infty$ can be obtained without carrying out any integral. Although it does not give a good quantitative match, it captures the high (low) density at large (small) angles.
By discarding all exponentials suppressed by $1/N$ from Eq.~(\ref{eq:complex_ratio_N}), factoring out terms containing $z$ from the $s$ and $t$ integrals, we obtain the prefactor,
\begin{equation}\label{eq:GinUE_Ninfty_polar}
    \varrho^{(N\to\infty)}_\mathrm{GinUE}(r,\theta)\approx\frac{12}{\pi}\frac{r^2(1+r^2-2r\cos\theta)}{(1+r^2)^5}\heav{1-r}.
\end{equation}

\begin{figure*}[tbp]
    \centering
    \includegraphics[width=0.99\textwidth]{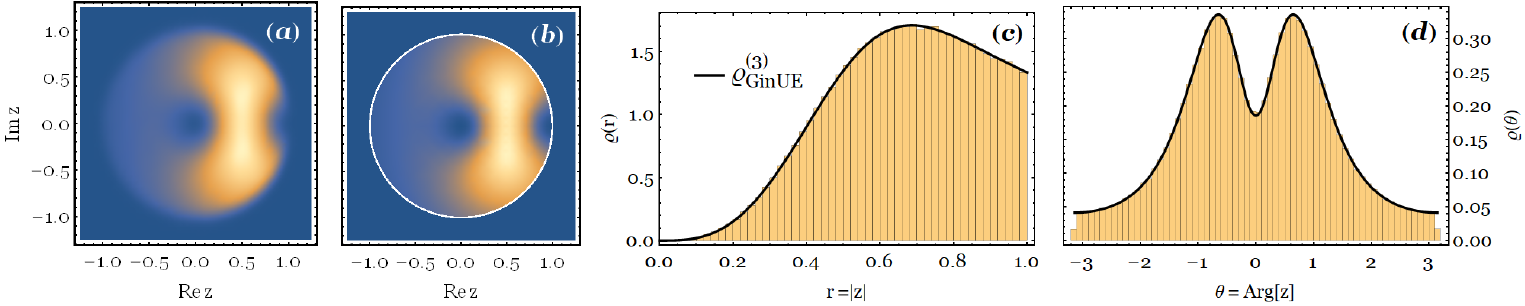}
    \caption{Comparison of exact diagonalization (ED) of GinUE-drawn matrices with analytical results, for $N=3$. $(a)$: ratio density from ED; $(b)$: exact distribution, Eq.~(\ref{eq:GinUE_N3}); $(c)$: histogram of absolute value of ratios from ED; $(d)$: histogram of argument of ratios from ED; black lines computed from Eq.~(\ref{eq:GinUE_N3}).}
    \label{fig:GinUE_TUE_N3}
\end{figure*}

\subsection{Toric Unitary Ensemble}
\label{subsection:ratio_nonhermitian_toric}

We now want to eliminate boundary effects from a complex spectrum by considering a two-dimensional analog of the circular ensembles. Recall that the circular ensembles has eigenvalues on the unit circle $\mathbb{S}^1$. A possible generalization would be to consider eigenvalues on the sphere $\mathbb{S}^2\subset\mathbb{R}^3$, which would be provided by the Spherical Unitary Ensemble (SUE)~\cite{forrester1992,brouwer1995,krishnapur2009,forrester2010,forrester2016}, of matrices $A^{-1}B$ with both $A$, $B$ GinUE matrices. However, while belonging to the same universality class as the GinUE, for the SUE the convergence to the large-$N$ limit is also quite slow. Instead, we consider eigenvalues on the two-dimensional (Clifford) torus $\mathbb{T}^2=\mathbb{S}^1\times\mathbb{S}^1\subset \mathbb{S}^3\subset\mathbb{R}^4$, which show a very fast convergence.

We parametrize the torus by two angles $\vartheta\in(-\pi,\pi]$, $\varphi\in(-\pi,\pi]$, with a generic point $P\in\mathbb{T}^2$ given by $P=(1/\sqrt{2})\left(\cos\vartheta,\sin\vartheta,\cos\varphi,\sin\varphi\right)$. In analogy with the CUE, we want to construct a flat joint eigenvalue distribution on $\mathbb{T}^2$, which we call the distribution of the toric unitary ensemble (TUE). Since the Clifford torus has no curvature, the distribution is simply given by the Vandermonde interaction, $P^{(N)}_\mathrm{TUE}(\vartheta_1,\dots,\vartheta_N;\varphi,\dots,\varphi_N)\propto \abs{\Delta_{\mathbb{T}^2}}^2$. $\Delta_{\mathbb{T}^2}$ is given by the distance between points in the embedding space parametrized by the eigenvalues. In other words, if $P_j\in\mathbb{T}^2$ is parametrized by the angles $(\vartheta_j,\varphi_j)$ then the Vandermonde interaction is $\abs{\Delta_{\mathbb{T}^2}}=\prod_{j<k}\norm{P_j-P_k}_{\mathbb{R}^4}$. One can check that, with this reasoning, the usual Vandermonde terms for the Gaussian, Ginibre, circular and spherical ensembles coincide with, respectively, $\abs{\Delta_{\mathbb{R}}}=\prod_{j<k}\norm{P_j-P_k}_{\mathbb{R}}$, $\abs{\Delta_{\mathbb{R}^2}}=\prod_{j<k}\norm{P_j-P_k}_{\mathbb{R}^2}$, $\abs{\Delta_{\mathbb{S}^1}}=\prod_{j<k}\norm{P_j-P_k}_{\mathbb{R}^2}$, $\abs{\Delta_{\mathbb{S}^2}}=\prod_{j<k}\norm{P_j-P_k}_{\mathbb{R}^3}$ (with $P_j$ in the respective embedding spaces). Using our parameterizantion of the torus and considering only $\beta=2$, the Vandermonde interaction reads $\abs{\Delta_{\mathbb{T}^2}}^2=\prod_{j<k}\left[2-\cos(\vartheta_j-\vartheta_k)-\cos(\varphi_j-\varphi_k)\right]$.

We can then write down the joint eigenvalue distribution for the TUE, given by Eq.~(\ref{eq:TUE_joint}),
\begin{equation*}
\begin{split}
    P^{(N)}_\mathrm{TUE}&(\vartheta_1,\dots,\vartheta_N;\varphi_1,\dots,\varphi_N)\propto \prod_{j<k}\left[2-\cos(\vartheta_j-\vartheta_k)-\cos(\varphi_j-\varphi_k)\right].
\end{split}
\end{equation*}

Having introduced the relevant joint eigenvalue distribution, the remaining procedure is straightforward. By rotational invariance in both factors $\mathbb{S}^1$, we can set $\vartheta_1=0$ and $\varphi_1=0$. The complex spacing ratio is, accordingly, $z=(\vartheta_2+i\varphi_2)/(\vartheta_3+i\varphi_3)$. If we then insert Eq.~(\ref{eq:TUE_joint}) into the general ratio distribution, Eq.~(\ref{eq:nonhermitian_ratio_general}), we obtain
\begin{equation}\label{eq:TUE_finite_N}
\begin{split}
    \varrho^{(N)}_\mathrm{TUE}(x,y)&\propto\int_{-\pi}^\pi\d s\d t \prod_{j=1}^{N-3}\d a_j \d b_j \heav{(a_j^2+b_j^2)-(s^2+t^2)}(s^2+t^2)^2\left[2-\cos s-\cos t\right]\\
    &\times\left[2-\cos(sx-ty)-\cos(t x+s y)\right]\left[2-\cos(s(x-1)-ty)-\cos(t (x-1)+s y)\right]\\
    &\times\prod_{j=1}^{N-3}\left[2-\cos a_j-\cos b_j\right]
    \left[2-\cos(s-a_j)-\cos(t-b_j)\right]\\
    &\times\left[2-\cos(sx-ty-a_j)-\cos(t x+s y-b_j)\right]\prod_{j<k}\left[2-\cos(a_j-a_k)-\cos(b_j-b_k)\right].
\end{split}
\end{equation}

For $N=3$, the double integral to be performed is given in Eq.~(\ref{eq:TUE_N3}),
\begin{equation*}
\begin{split}
    \varrho^{(3)}_\mathrm{TUE}(x,y)&\propto\int_{-\pi}^\pi\d s\d t (s^2+t^2)^2\left[2-\cos s-\cos t\right]\left[2-\cos(sx-ty)-\cos(t x+s y)\right]\\
    &\times\left[2-\cos(s(x-1)-ty)-\cos(t (x-1)+s y)\right].
\end{split}
\end{equation*}
The integral of Eq.~(\ref{eq:TUE_N3}) and its generalizations for $N=4,5,\dots$ can be numerically integrated (the analytic expression is far too involved to be useful) and describe very well the large-$N$ asymptotics of the GinUE universality class, see Figs.~\ref{fig:analytics_GinUE}-$(e)$--$(h)$.

\end{widetext}

\bibliography{bibfile}

\end{document}